\documentstyle[12pt]{article}
\font\titlefont=cmbx10 scaled \magstep5

\begin{document}

\input{epsf}

\begin{flushright}
\vspace*{-2cm}
 gr-qc/9707062  \\  TUTP-97-9 \\ July 30, 1997
\vspace*{2cm}
\end{flushright}

\begin{center}
{\titlefont QUANTUM FIELD THEORY }\\
\vspace{0.5cm}
{\titlefont IN CURVED SPACETIME\footnote{Lectures given at the IX Jorge Andr\'e
Swieca Summer School, Campos dos Jord\~ao, SP, Brazil, February 1997,
and at Soochow University, Taipei, Taiwan, June 1997. To be published in
the proceedings of the Swieca School.}}\\
\vskip .6in
L.H. Ford \\
\vskip .3in
Institute of Cosmology\\
Department of Physics and Astronomy\\
Tufts University\\
Medford, Massachusetts 02155\\
email: ford@cosmos2.phy.tufts.edu\\
\end{center}

\vskip 0.6in
\baselineskip=13pt

\centerline{\bf OUTLINE}
\vskip .3in
     These lectures will deal with selected aspects of quantum field theory in 
curved spacetime \cite{1993}. The basic outline of this series of lectures 
will be as follows:
\begin{itemize}
\item {Lecture 1.} Quantization of fields on a curved background, particle
creation by gravitational fields, particle creation in an expanding universe;
moving mirror radiation.
\item {Lecture 2.} The Hawking effect - particle creation by black holes.
\item {Lecture 3.} Ultraviolet and infrared divergences, renormalization
of the expectation value of the stress tensor; global symmetry breaking in
curved spacetime.
\item {Lecture 4.} Negative energy in quantum field theory, its gravitational
effects, and inequalities which limit negative energy densities and fluxes.
\item {Lecture 5.} The semiclassical theory of gravity and its limitations,
breakdown of this theory due to metric fluctuations, lightcone fluctuations.
\end{itemize}

\newpage

\baselineskip=14pt
\section{BASIC FORMALISM AND \hfil\break PARTICLE CREATION}
\subsection{Second Quantization in Curved Space}

\setcounter{equation}{0}
\renewcommand{\theequation}{1.\arabic{equation}}

      There are four basic ingredients in the construction of a quantum field
theory. These are
\begin{itemize}
\item The Lagrangian, or equivalently, the equation of motion of the
classical theory.
\item A quantization procedure, such as canonical quantization or the
path integral approach.
\item The characterization of the quantum states.
\item The physical interpretation of the states and of the observables.
\end{itemize}

In flat spacetime, Lorentz invariance plays an important role in each of 
these steps. For example, it is a guide which generally allows us to identify 
a unique vacuum state for the theory. However, in curved spacetime, we
do not have Lorentz symmetry. This is not a crucial problem in the first two
steps listed above. The formulation of a classical field theory and its
formal quantization may be carried through in an arbitrary spacetime. The
real differences between flat space and curved space arise in the latter
two steps. In general, there does not exist a unique vacuum state in a curved
spacetime. As a result, the concept of particles becomes ambiguous, and
the problem of the physical interpretation becomes much more difficult.

    The best way to discuss these issues in more detail is in the context 
of a particular model theory. Let us consider a real, massive scalar field
for which the Lagrangian density is  
\begin{equation}
{\cal L}={1\over 2}(\partial _\alpha \varphi \partial ^\alpha \varphi -
          m^2 \varphi^2- \xi R \varphi^2).
\end{equation}
(We adopt the sign conventions of Birrell and Davies \cite{BD}, which are
the $(- - -)$ conventions in the notation of Misner, Thorne, and 
Wheeler \cite{MTW}. In particular, the metric signature will be $(+ - - -)$.
Unless otherwise noted, units in which $G=c=\hbar=1$ are used.) 
The corresponding wave equation is
\begin{equation}
\Box \varphi +m^2\varphi +\xi R\varphi =0. \label{eq:KG}
\end{equation}
Here $R$ is the scalar curvature, and $\xi$ is a new coupling constant.
There are two popular choices for $\xi$: minimal coupling ($\xi =0$) and
conformal coupling ($\xi = {1\over 6}$). The former leads to the simplest
equation of motion, whereas the latter leads to a theory which is conformally
invariant in four dimensions in the massless limit. 
For our purposes, we need not settle
this issue, but rather regard $\xi$ on the same footing as $m$, as a parameter
which specifies our theory. Note that here $\Box$ denotes the generally
covariant d'Alembertian operator, $\Box = \nabla_\mu \,\nabla^\mu $.

    A useful concept is that of the {\it inner product} of a pair of solutions
of the generally covariant Klein-Gordon equation, Eq.(\ref{eq:KG}). 
It is defined by
\begin{equation}
(f_1,f_2)=i\int (f^*_2 \, {\mathop{\partial_\mu}\limits^\leftrightarrow } \,f_1)
d\Sigma^\mu,
\end{equation}
where $d\Sigma^\mu = d\Sigma\, n^\mu$, with $d\Sigma$ being the volume element
in a given spacelike hypersurface, and $n^\mu$ being the timelike unit vector
normal to this hypersurface. The crucial property of the inner product
is that it is independent of the choice of hypersurface. That is, if
$\Sigma_1$ and $\Sigma_2$ are two different, non-intersecting hypersurfaces,
then 
\begin{equation}
(f_1,f_2)_{\Sigma_1} = (f_1,f_2)_{\Sigma_2}. \label{eq:inner}
\end{equation}
The proof of this property is straightforward. We assume that $f_1$ and
$f_2$ are both solutions of Eq. (\ref{eq:KG}). Furthermore, 
if the space is such that
the hypersurfaces are non-compact, we assume that these functions vanish at
spatial infinity. Let $V$ be the four-volume bounded by  $\Sigma_1$ and 
$\Sigma_2$, and, if necessary, time-like boundaries on which $f_1 = f_2=0$.
Then we may write
\begin{equation}
(f_1,f_2)_{\Sigma_2} - (f_1,f_2)_{\Sigma_1} =
i\oint_{\partial V} (f^*_2 \,
{\mathop{\partial_\mu}\limits^\leftrightarrow } \,f_1)
d\Sigma^\mu =
i\int_{V}\nabla_\mu(f^*_2 \,
{\mathop{\partial_\mu}\limits^\leftrightarrow } \,f_1) dV,
\end{equation}
where the last step follows from the four dimensional version of Gauss' law,
and $dV$ is the four dimensional volume element.
However, we may write this integrand as
\begin{eqnarray}
\nabla_\mu(f^*_2 \,{\mathop{\partial_\mu}\limits^\leftrightarrow } \,f_1) 
&=&\nabla_\mu(f^*_2 \partial_\mu \,f_1  -f_1 \partial_\mu \, f^*_2) =
f^*_2 \Box f_1 - f_1 \Box f^*_2 \nonumber \\
&=&-f^*_2 (m^2 +\xi R) f_1 + f_1 (m^2 +\xi R) f^*_2 =0.
\end{eqnarray}
Thus Eq. (\ref{eq:inner}) is proven. 

     The quantization of a scalar field in a curved spacetime may be 
carried out by canonical methods. Choose a foliation of the spacetime
into spacelike hypersurfaces. Let $\Sigma$ be a particular hypersurface
with unit normal vector $n^\mu$ labelled by a constant value of the time
coordinate $t$. The derivative of $\varphi$ in the normal direction is
$\dot \varphi = n^\mu\, \partial_\mu \varphi$, and the canonical momentum
is defined by
\begin{equation}
\pi = {{\delta \cal L}\over {\delta \dot \varphi}}\quad .
\end{equation}
We impose the canonical commutation relation
\begin{equation}
[\varphi({\bf x},t), \pi({\bf x}',t)] = i\delta({\bf x},{\bf x}'),
\end{equation}
where $\delta({\bf x},{\bf x'})$ is a delta function in the hypersurface
with the property that 
\begin{equation}
\int \delta({\bf x},{\bf x}') d\Sigma =1 \, .
\end{equation}

     Let $\{ f_j \}$ be a complete set of positive norm solutions of
Eq. (\ref{eq:KG}). 
Then $\{ f^*_j \}$ will be a complete set of negative norm solutions,
and $\{ f_j, f^*_j \}$ form a complete set of solutions of the wave equation
in terms of which we may expand an arbitrary solution. Write the field
operator $\varphi$ as a sum of annihilation and creation operators:
\begin{equation}
\varphi = \sum_j (a_j f_j + a^\dagger_j f^*_j),
\end{equation}
where $[a_j, a^\dagger_{j'}] = \delta_{j,j'}$. This expansion defines a
vacuum state $|0\rangle$ such that $a_j |0\rangle=0$. In flat spacetime,
we take our positive norm solutions to be positive frequency solutions,
$f_j \propto e^{-i\omega t}$. Regardless of the Lorentz frame in which
$t$ is the time coordinate, this procedure defines the same, unique
Minkowski vacuum state.

     In curved spacetime, the situation is quite different. There is, 
in general,
no unique choice of the $\{ f_j \}$, and hence no unique notion of the
vacuum state. This means that we cannot identify what constitutes a state
without particle content, and the notion of ``particle'' becomes
ambiguous. One possible resolution of this difficulty is to choose some
quantities other than particle content to label quantum states. Possible
choices might include local expectation values \cite{algebraic}, 
such as $\langle \varphi
\rangle$, $\langle \varphi^2 \rangle$, etc.  In the particular case of
an asymptotically flat spacetime, we might use the particle content in
an asymptotic region. Even this characterization is not unique. However,
this non-uniqueness is an essential feature of the theory with physical
consequences, namely the phenomenon of particle creation, which we will
now discuss.

\subsection{Particle Creation by Gravitational Fields}

    Let us consider a spacetime which is asymptotically flat in the past
and in the future, but which is non-flat in the intermediate region.
Let $\{ f_j \}$ be positive frequency solutions in the past (the 
``in-region''), and let $\{ F_j \}$ be positive frequency solutions in 
the future (the ``out-region'').  We may choose these sets of solutions
to be orthonormal, so that
\begin{eqnarray}
&(f_j,f_{j'})= (F_{j},F_{j'})=\delta _{jj'}& \nonumber \\
&(f_j^*,f_{j'}^*)= (F_{j}^*,F_{j'}^*)= -\delta _{jj'}& \nonumber \\
&(f_j,f_{j'}^*)= (F_{j},F_{j'}^*)= 0.&  \label{eq:ortho}
\end{eqnarray}
Although these functions are defined by their asymptotic properties in
different regions, they are solutions of the wave equation everywhere
in the spacetime. We may expand the in-modes in terms of the out-modes:
\begin{equation}
f_j=\sum\limits_k {(\alpha _{jk}}F_k + \beta _{jk}F_k^*).
\end{equation}
Inserting this expansion into the orthogonality relations, 
Eq. (\ref{eq:ortho}), leads to the conditions
\begin{equation}
\sum\limits_k {(\alpha _{jk}\alpha _{j'k}^*-\beta _{jk}\beta_{j'k}^*)
= \delta _{jj'}}, \label{eq:alphabeta}
\end{equation}
and
\begin{equation}
 \sum\limits_k (\alpha _{jk}\alpha _{j'k}-\beta _{jk}\beta _{j'k})=0.
\end{equation}
The inverse expansion is 
\begin{equation}
F_k=\sum\limits_j {(\alpha _{jk}^*}f_j - \beta _{jk}f_j^*).
\end{equation}

The field operator, $\varphi$, may be expanded in terms of either the
$\{ f_j \}$ or the $\{ F_j \}$:
\begin{equation}
  \varphi = \sum\limits_j (a_j f_j + a_j^\dagger f_j^*)
  =\sum\limits_j (b_j F_j + b_j^\dagger F_j^*).
\end{equation}
The $a_j$ and $a_j^\dagger$ are annihilation and creation operators,
respectively, in the in-region, whereas the $b_j$ and $b_j^\dagger$
are the corresponding operators for the out-region. The in-vacuum state
is defined by $a_j|0\rangle_{in}=0, \; \forall j,$ and describes the
situation when no particles are present initially. The out-vacuum state
is defined by $b_j|0\rangle_{out}=0, \; \forall j,$ and describes the
situation when no particles are present at late times. Noting that
$a_j = (\varphi,f_j)$ and $b_j = (\varphi,F_j)$, we may expand the
two sets of creation and annihilation operator in terms of one another
as 
\begin{equation}
a_j=\sum\limits_k (\alpha _{jk}^*b_k-\beta _{jk}^* b_k^\dagger),
                                      \label{eq:Bogo1}
\end{equation}
or 
\begin{equation}
b_k=\sum\limits_j (\alpha _{jk} a_j + \beta _{jk}^* a _j^\dagger).
                                       \label{eq:Bogo2}
\end{equation}
This is a Bogolubov transformation, and the $\alpha_{jk}$ and
$\beta_{jk}$ are called the Bogolubov coefficients.

   Now we are ready to describe the physical phenomenon of particle
creation by a time-dependent gravitational field. Let us assume that
no particle were present before the gravitational field is turned on.
If the Heisenberg picture is adopted to describe the quantum dynamics,
then $|0\rangle_{in}$ is the state of the system for all time. However,
the physical number operator which counts particles in the out-region
is $N_k = b_k^\dagger b_k$. Thus the mean number of particles created
into mode $k$ is 
\begin{equation}
\langle N_k \rangle = {}_{in}\langle 0|b_k^\dagger b_k |0\rangle_{in}
       = \sum\limits_j |{\beta _{jk}}|^2.
\end{equation}
If any of the $\beta_{jk}$ coefficients are non-zero, i.e. if
any mixing of positive and negative frequency solutions occurs, then
particles are created by the gravitational field. 

      The most straightforward application of the concepts developed above
is to particle creation by an expanding universe. This phenomenon was
first hinted at in the work of Schr\"odinger \cite{Schrodinger}, 
but was first carefully investigated by Parker \cite{Parker}. 
Let us restrict our attention to the case of a
spatially flat Robertson-Walker universe, for which the metric may be
written as
\begin{equation}
ds^2 = dt^2 -a^2(t) d{\bf x}^2 = a^2(\eta)\,
       \bigl(d\eta^2 - d{\bf x}^2\bigr),
\end{equation}
where $a$ is the scale factor. We may use either the comoving time $t$
or the conformal time $\eta$, but the solutions of the wave equation are
simpler in terms of the latter. The positive norm solutions of 
Eq. (\ref{eq:KG}) in this metric may be taken to be 
\begin{equation}
f_{\bf k}({\bf x},\eta)={{e^{i{\bf k\cdot x}}} \over 
{a(\eta)\sqrt {(2\pi)^3}}}\,\, \chi_k(\eta), 
\end{equation}
where $\chi_k(\eta)$ satisfies 
\begin{equation}
{{d^2\chi_k} \over {d\eta^2}} + [ {k^2- V(\eta)} ]\chi_{k} =0 \, ,
                                                 \label{eq:chieq}
\end{equation}
with
\begin{equation}
V(\eta) \equiv -a^2(\eta) 
        \biggl[m^2 + \Bigl( \xi -{1 \over 6}\Bigr) R(\eta)\biggr].
\end{equation}
The norm of $f_{\bf k}$ being equal to one is equivalent to the Wronskian 
condition
\begin{equation}
\chi_k {{d \chi_k^*} \over {d\eta}} - \chi_k^* {{d \chi_k} \over {d\eta}} =i.
\end{equation}

Let us consider the idealized situation in which the universe is static 
both in the past and in the future. In this case, we have the necessary
asymptotically flat regions needed to define in and out vacua. Let us
make the further simplification that the field is massless, $m=0$.   
We have chosen modes which are pure positive frequency in the past,
the in-modes:
\begin{equation}
\chi_k(\eta) \sim \chi_k^{(in)}(\eta) = {{e^{-i\omega \eta}}\over
{\sqrt{2\omega}}}, \qquad \eta \rightarrow -\infty. \label{eq:chipast}
\end{equation}
Their form in the future is
\begin{equation}
\chi_k(\eta) \sim \chi_k^{(out)}(\eta) = {1\over {\sqrt{2\omega}}}
\bigl(\alpha_k {e^{-i\omega \eta}} + \beta_k {e^{i\omega \eta}}\bigr), 
\qquad \eta \rightarrow \infty,    \label{eq:chifuture}
\end{equation}
where the coefficients $\alpha_k$ and $\beta_k$ are determined by solving 
Eq. (\ref{eq:chieq}) for a given $a(\eta)$. 
They are related to the Bogolubov coefficients
by $\alpha_{\bf k k'} = \alpha_k \delta_{\bf k k'}$ and
$\beta_{\bf k k'} = \beta_k \delta_{\bf k, -k'}$. Thus the number density 
of created particles per unit proper volume  is given at late times by
\begin{equation}
N = {1 \over {(2\pi a)^3}} \int d^3k\, |\beta_k|^2,
\end{equation}
and their energy density by
\begin{equation}
\rho = {1 \over {(2\pi a)^3 a}} \int d^3k\, \omega |\beta_k|^2.
\end{equation}
These formulas are to be understood to hold in the asymptotic region where 
the particle creation has effectively stopped, and $a$ is the scale factor
in that region. Thus the number of particles per unit proper volume is
proportional to $a^{-3}$, and energy per particle is proportional to 
$a^{-1}$. Note that we are here discussing massless particles 
whose wavelength is sufficiently short that they redshift as would conformally
invariant massless particles.

    Unfortunately, it is difficult to solve Eq. (\ref{eq:chieq}) for the mode 
functions in all but the simplest examples. However, there is a perturbative
method, developed by Zeldovich and Starobinsky \cite{ZS} and by Birrell and
Davies \cite{BD2}, which is often useful. The first step is to rewrite the
differential equation for $\chi_k$ as an integral equation:
\begin{equation}
\chi_k(\eta) = \chi_k^{(in)}(\eta) + \omega^{-1} \int_{-\infty}^\eta
             V(\eta')\;\sin \, \omega(\eta-\eta')\;\chi_k(\eta') d\eta'.
\end{equation}
This integral equation is equivalent to Eq. (\ref{eq:chieq}) 
plus the boundary condition Eq. (\ref{eq:chipast}). 
We now wish to assume that $V$ is sufficiently small that we
may iterate this integral equation to lowest order by replacing
$\chi_k(\eta')$ by $\chi_k^{(in)}(\eta')$ in the integrand. If we
compare the resulting formula with Eq. (\ref{eq:chifuture}),  
the Bogolubov coefficients may be read off:
\begin{equation}
\alpha_k \approx 1 + {i \over {2\omega}} \int_{-\infty}^\infty
             V(\eta)\, d\eta,
\end{equation}
and
\begin{equation}
\beta_k \approx - {i \over {2\omega}} \int_{-\infty}^\infty
            e^{-2i\omega \eta} V(\eta)\, d\eta.
\end{equation}
Let us restrict our attention to the case where $m=0$. In this case, the
mean number density becomes
\begin{equation}
N= {{(\xi -{1\over 6})^2}\over {16\pi a^3}} \int_{-\infty}^\infty
                        a^4(\eta)\,R^2(\eta)\,d\eta,
\end{equation}
and the energy density becomes
\begin{eqnarray}
\rho= - {{(\xi -{1\over 6})^2}\over {32\pi^2 a^4}} \int_{-\infty}^\infty d\eta_1
    \int_{-\infty}^\infty d\eta_2 &{}& \!\!\!\!\!\!\!\!\! 
    \biggl\{ \ln(|\eta_1-\eta_2|\mu)
    {d\over {d\eta_1}}\Bigl[a^2(\eta_1)\,R(\eta_1)\Bigr] \nonumber \\
    &\times &{d\over {d\eta_2}}\Bigl[a^2(\eta_2)\,R(\eta_2)\Bigr] \biggr\}\,.
\end{eqnarray}
Here $\mu$ is an arbitrary quantity with the dimensions of mass; $\rho$
is independent of $\mu$ provided that $a^2(\eta)\,R(\eta) \rightarrow \infty$
as $\eta \rightarrow \pm \infty$.
The approximation which  is being used here amounts to perturbation around
the conformally invariant theory in powers of $(\xi -{1\over 6})$.

    As an application of these formulas, let us consider particle creation
at the end of an inflationary expansion. A typical inflationary scenario
involves the universe making a transition from deSitter space to a radiation
dominated Robertson-Walker universe on a relatively short time scale. It
is usually assumed that there is a mechanism for creating matter via
particle interactions. However, there will also be at least some matter
generated by gravitational particle creation. We may use the above results
to make some order of magnitude estimates for massless, non-conformal
scalar particles \cite{F87}. Let $\Delta t$ be the duration of the transition in
co-moving time. The scalar curvature in the deSitter phase is given by
$R=12H^2$, where $H^{-1}$ is the e-folding time of the inflationary
expansion. The scalar curvature drops to zero in the radiation dominated
phase. If we assume that the transition occurs rapidly so that
$\Delta t \ll H^{-1}$, then we have approximately that
\begin{equation}
N \approx {{(\xi -{1\over 6})^2}\over {12\pi a^3}} H^3, 
\end{equation}
and that
\begin{equation}
\rho \approx {{(\xi -{1\over 6})^2 H^4}\over {8\pi^2 a^4}} 
            \;  \ln\biggl({1 \over {H\Delta t}}\biggr).
\end{equation}
Comparison of these two results indicates that the mean energy of
the created particles is of order $H \ln[(H\Delta t)^{-1}]$. Note in the limit
$\Delta t \rightarrow 0$, that $N$ is finite, but $\rho$ is unbounded.
The vacuum energy density, $\rho_V$, which drives the expansion, is related to
$H$ by the Einstein equation:
\begin{equation}
H^2 = {{8\pi\rho_V}\over {3\sqrt{\rho_{Pl}}}} \, ,  
\end{equation}
where $\rho_{Pl} \approx \bigl( 10^{19} GeV\bigr)^4$ is the Planck density.
We can express our estimate for the energy density of the created particles
just after the end of inflation as
\begin{equation}
\rho \approx (1-6\xi)^2 {{\rho_V^2}\over {\rho_{Pl}}}.
\end{equation}
If, for example, we were to take $\rho_V \approx \bigl( 10^{15}\, GeV\bigr)^4$,
which is a typical value for inflation at the GUT (Grand Unified Theory) 
scale, then we obtain the estimate 
\begin{equation}
\rho \approx (1-6\xi)^2 \bigl( 10^{11}\, GeV\bigr)^4.
\end{equation}
This energy density is much less than $\rho_V$, and would hence be negligible 
if there is efficient reheating. However, if other reheating mechanisms are
not efficient, then particle creation by the gravitational field could
play a significant role in cosmological evolution. 

\subsection{Particle Creation by Moving Mirrors}
\label{sec:mirror}

    A simple example of quantum particle creation was given by Fulling and 
Davies \cite{FD76, DF77}. This consists of a moving mirror in two-dimensional
spacetime coupled to a massless scalar field, $\varphi$. The field is assumed
to satisfy a boundary condition on the worldline of the mirror, such as
$\varphi = 0$. For a given mirror trajectory, it is possible to construct
exact solutions of the wave equation which satisfy this boundary condition.
Let $v=t+x$ and $u=t-x$ be null coordinates which are constant upon null rays
moving to the left and to the right, respectively. A null ray of fixed $v$ 
which reflects off of the mirror becomes a ray of fixed $u$. The relation 
between the values of $v$ and of $u$ is a function determined by the mirror's
trajectory (See Figure 1.). Let
\begin{equation}
v = G(u) \,,
\end{equation}
or, equivalently, 
\begin{equation}
u = g(v) = G^{-1}(v) \,.
\end{equation}
The mode functions which satisfy the massless wave equation and which vanish
on the worldline of the mirror are
\begin{equation}
f_k(x) = \frac{1}{\sqrt{4 \pi \omega}}\, 
         \Bigl( e^{-i\omega v} - e^{-i\omega G(u)} \Bigr) \,.
\end{equation}
The incoming positive frequency wave, $e^{-i\omega v}$, is reflected from the
mirror and becomes an outgoing wave, $e^{-i\omega G(u)}$, which is a 
superposition of positive frequency  ($e^{-i\omega u}$) and negative
frequency ($e^{i\omega u}$) parts. 

\begin{figure}
\begin{center}
\leavevmode\epsfysize=7.5cm\epsffile{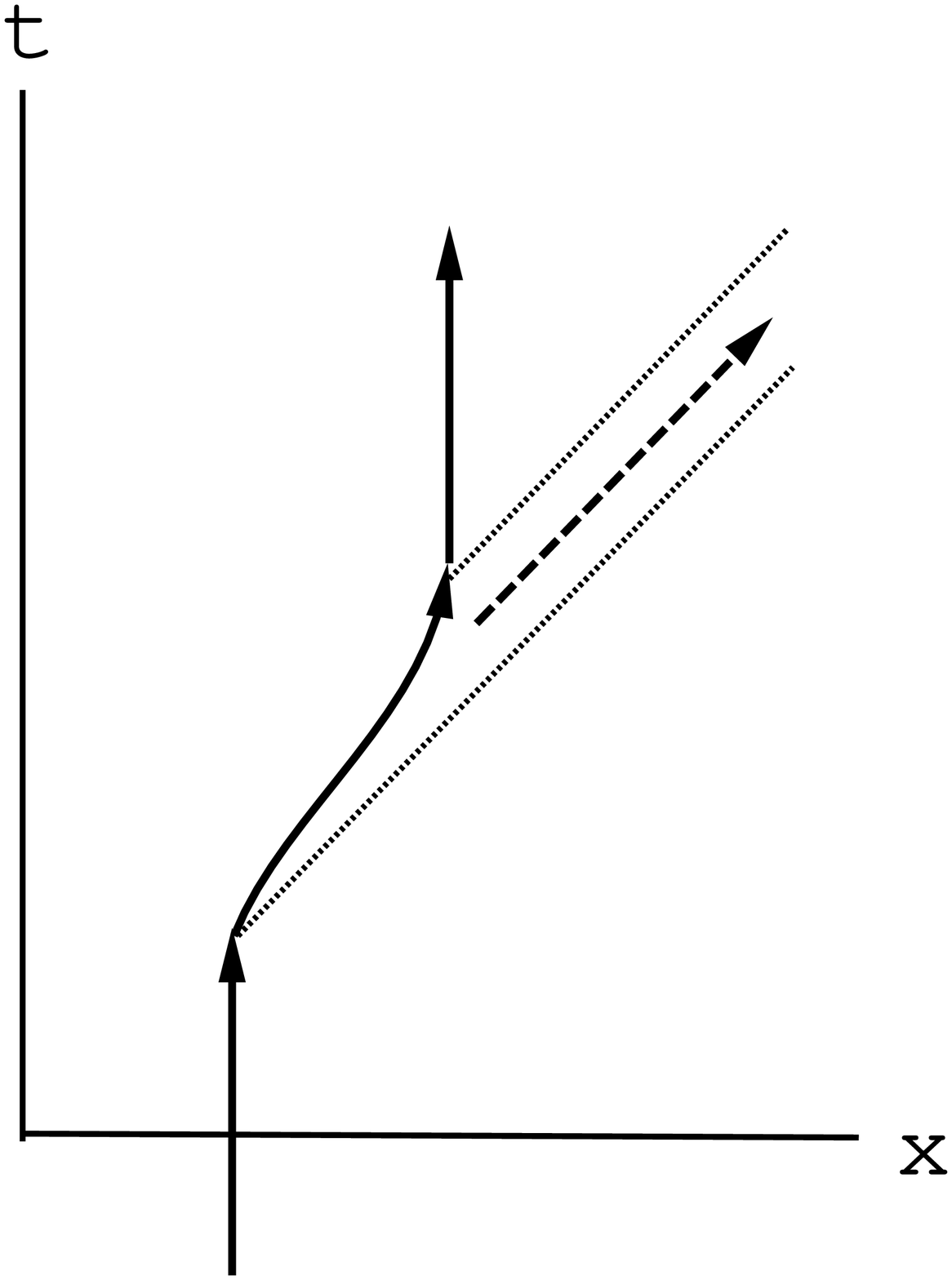}
\label{Figure 1}
\end{center}
\begin{caption}[]

A moving mirror in two-dimensional spacetime accelerates for a finite period of
time. The quantum radiation emitted to the right of the mirror propagates in
the spacetime region between the dotted lines. There is also radiation emitted
to the left which is not shown.
\end{caption}
\end{figure}

   Fulling and Davies \cite{FD76} show that the flux of energy radiated to the 
right is
\begin{equation}
F(u) = \langle T^{xt} \rangle = 
\frac{1}{48 \pi} 
\biggl[ 3\biggl(\frac{G''}{G'}\biggr)^2 -2\biggl(\frac{G'''}{G'}\biggr)\biggr]
                     \,.  \label{eq:mirror}
\end{equation}
This flux may also be expressed in terms of the instantaneous mirror velocity
$v(t)$ as 
\begin{equation}
F = - \frac{(1-v^2)^{1/2}}{12 \pi (1-v)^2} \, \frac{d}{d t} \,
            \left[ \frac{\dot v}{(1-v^2)^{3/2}} \right] \, ,  \label{eq:mirror2}
\end{equation}
 In the nonrelativistic limit, 
\begin{equation}
F \approx - \frac{\ddot v}{12 \pi} \, .  \label{eq:mirror3}
\end{equation}
 Note that $F$ may be either positive or 
negative. In the latter case, one has an example of the negative energy in
quantum field theory that will be the topic of  Lecture \ref{sec:negen}.

   Unfortunately, the simple solution for the moving mirror radiation of a
massless field in two-dimensional spacetime depends upon the special conformal
properties of this case and does not generalize to massive fields or to 
four-dimensional spacetime. In the four-dimensional case, there are exact
solutions available for special trajectories \cite{CD77,FS79}, and approximate
solutions for general trajectories \cite{FV82a}, but no general, exact solutions.
However, the technique of mapping between ingoing and outgoing rays is crucial
in the derivation of particle creation by black holes, which is the topic of
the next lecture.

\vspace{1cm}

\section{THE HAWKING EFFECT }
\setcounter{equation}{0}
\renewcommand{\theequation}{2.\arabic{equation}}

    In this lecture, we will apply the notions of particle creation by 
gravitational fields to black hole spacetimes. This leads to the Hawking
effect \cite{Hawking74,Hawking}, the process by which black holes 
emit a thermal spectrum of particles.
For the sake of definiteness, we will concentrate on the case of a massless,
scalar field in the Schwarzschild spacetime, but the basic ideas may be 
applied to any quantum field in a general black hole spacetime.
For the most part, we will follow the original derivation given by 
Hawking \cite{Hawking}. We imagine that the black hole was formed at some
time in the past by gravitational collapse. The spacetime of a collapsing
star is illustrated in Fig. 2.  This is not only physically
reasonable, but also avoids the issue of boundary conditions on the past
horizon which would arise if we were to consider the full Schwarzschild
spacetime.

\begin{figure}
\begin{center}
\leavevmode\epsfysize=12cm\epsffile{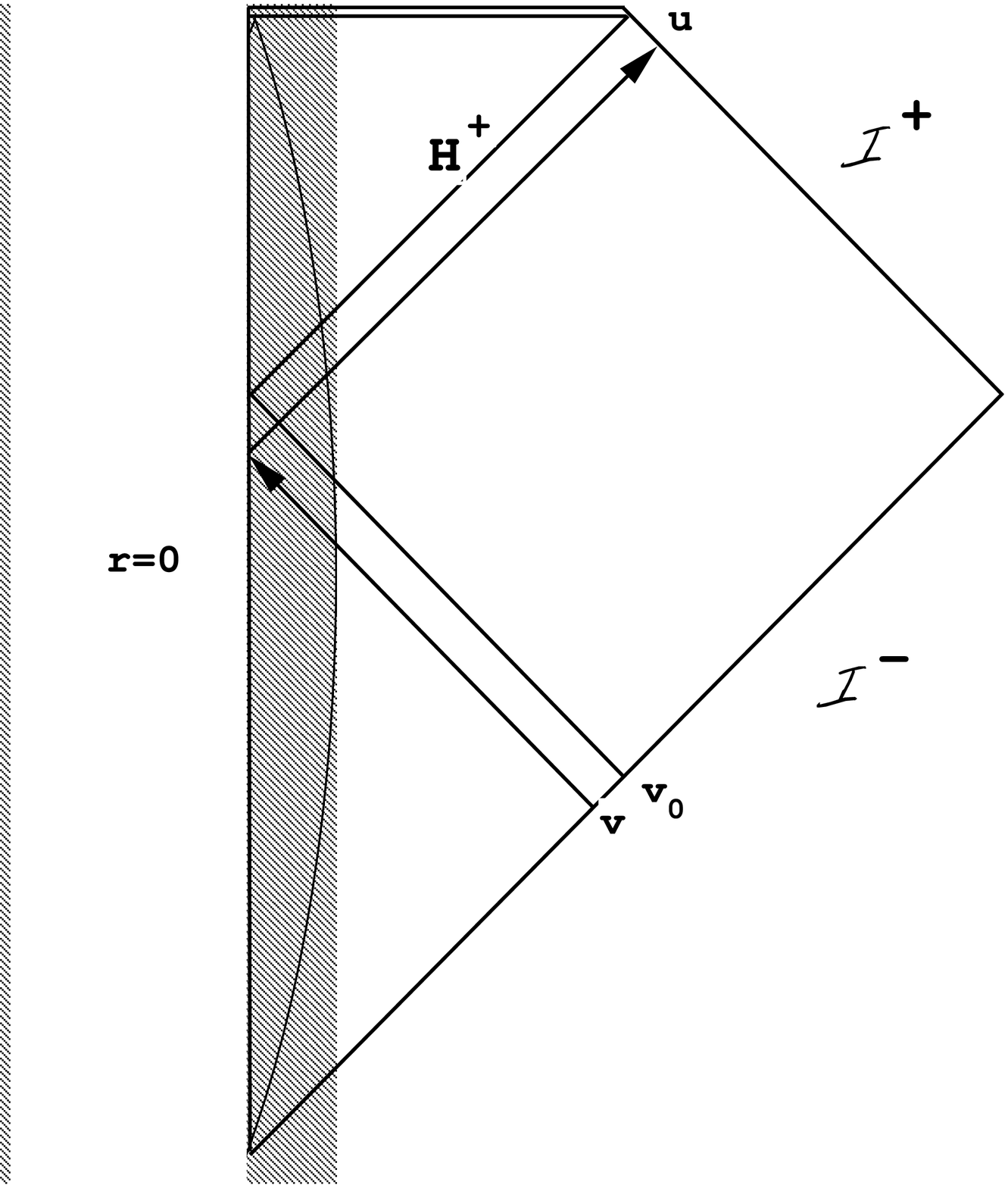}
\label{Figure 2}
\end{center}
\begin{caption}[]

The Penrose diagram for the spacetime of a black hole
formed by gravitational collapse. The shaded region is the interior of the
collapsing body, the $r=0$ line on the left is worldline of the center 
of this body, the $r=0$ line at the top of the diagram is the curvature
singularity, and $H^{+}$ is the future event horizon. An ingoing light
ray with $v < v_0$ from ${\cal I^{-}}$ passes through the body and escapes
to ${\cal I^{+}}$ as a $u = constant$ light ray. Ingoing rays with $v > v_0$
do not escape and eventually reach the singularity.
\end{caption}
\end{figure}

  Let us assume that 
no scalar particles were present before the collapse began. In this case,
the quantum state is the in-vacuum: $|\psi\rangle = |0\rangle_{in}.$
The in-modes, $f_{\omega\ell m}$, are pure positive frequency on
${\cal I^{-}}$, so $f_{\omega\ell m} \sim e^{-i\omega v}$ as $v \rightarrow
-\infty$, where $v=t+r^*$ is the advanced time coordinate. Similarly, the
out-modes, $F_{\omega\ell m}$, are pure positive frequency on
${\cal I^{+}}$, so $F_{\omega\ell m} \sim e^{-i\omega u}$ as $u \rightarrow
\infty$, where $u=t-r^*$ is the retarded time coordinate. As before, we
need to find the relation between these two sets of modes in order to 
calculate the Bogolubov coefficients and determine the particle creation.
Fortunately, it is not necessary to explicitly solve the wave equation
for the modes everywhere in order to determine the Bogolubov coefficients.
We are primarily interested in particle emission at late times (long after
the collapse occurs). This is dominated by modes which left ${\cal I^{-}}$
with very high frequency, propagated through the collapsing body just
before the horizon formed, and then underwent a large redshift on the 
way out to ${\cal I^{+}}$. Because these modes had an extremely high frequency
during their passage through the body, we may describe their propagation
by use of {\it geometrical optics}. 

     A $v= constant$ ingoing ray passes through the body and emerges as
a $u= constant$ outgoing ray, where $u =g(v)$ or equivalently, 
$v= g^{-1}(u) \equiv G(u)$. The geometrical optics approximation leads
to the following asymptotic forms for the modes:  
\begin{equation}
f_{\omega\ell m} \sim \frac{Y_{\ell m}(\theta,\phi)}{\sqrt{4\pi\omega}\,r}
 \times \cases{e^{-i\omega v}, &on ${\cal I^{-}}$ \cr
                             e^{-i\omega G(u)}, &on ${\cal I^{+}}$ \cr}
\end{equation}
and
\begin{equation}
F_{\omega\ell m} \sim \frac{Y_{\ell m}(\theta,\phi)}{\sqrt{4\pi\omega}\,r}
 \times \cases{e^{-i\omega u}, &on ${\cal I^{+}}$ \cr
                             e^{-i\omega g(v)}, &on ${\cal I^{-}}$\,, \cr}
                                   \label{eq:asymp}
\end{equation}
where $Y_{\ell m}(\theta,\phi)$ is a spherical harmonic. 
Hawking \cite{Hawking} gives a general ray-tracing argument which leads
to the result that 
\begin{equation}
u=g(v) = -4M\, \ln\biggl({{v_0 -v}\over C}\biggr), \label{eq:uofv}
\end{equation}
or
\begin{equation}
v=G(u) = v_0 -C e^{-u/4M}, \label{eq:vofu}
\end{equation}
where $M$ is the black hole mass, $C$ is a constant, and $v_0$ is the limiting
value of $v$ for rays which pass through the body before the horizon forms.

   We will derive this result for the explicit case of a thin shell. The
spacetime inside the shell is flat and may be described by the metric
\begin{equation}
ds^2 = dT^2 -dr^2 -r^2\,d\Omega^2.
\end{equation}
Thus, in the interior region, $V=T+r$ and $U=T-r$ are null coordinates which 
are constant on ingoing and on outgoing rays, respectively. The exterior of
the shell is a Schwarzschild spacetime with the metric
\begin{equation}
ds^2 = \Bigl(1 -{{2M}\over r}\Bigr) dt^2 -
        \Bigl(1 -{{2M}\over r}\Bigr)^{-1} dr^2 -r^2\,d\Omega^2.
\end{equation}
As noted above, the null coordinates here are $v=t+r^*$ and $u=t-r^*$,
where 
\begin{equation}
r^* = r +2M\, \ln\biggl({{r -2M}\over {2M}}\biggr)
\end{equation}
is the ``tortoise coordinate''. Let $r=R(t)$ describe the history of
the shell. The metric in this three dimensional hypersurface must
be the same as seen from both sides of the shell. (The intrinsic geometry
must match.) This leads to the condition
\begin{equation}
1- \biggl({{dR}\over {dT}}\biggr)^2 = 
 \biggl({{R -2M}\over {R}}\biggr)\biggl({{dt}\over {dT}}\biggr)^2
- \biggl({{R -2M}\over {R}}\biggr)^{-1}\biggl({{dR}\over {dT}}\biggr)^2.
                                          \label{eq:intrinsic}
\end{equation}
There is a second junction condition, that the extrinsic curvatures
of each side of this hypersurface match \cite{MTW2}. This leads to the equation
which determines $R(t)$ in terms of the stress-energy in the shell.
For our purposes, this equation is not needed, and we may assume an
arbitrary $R(t)$. 

\begin{figure}
\begin{center}
\leavevmode\epsfysize=6cm\epsffile{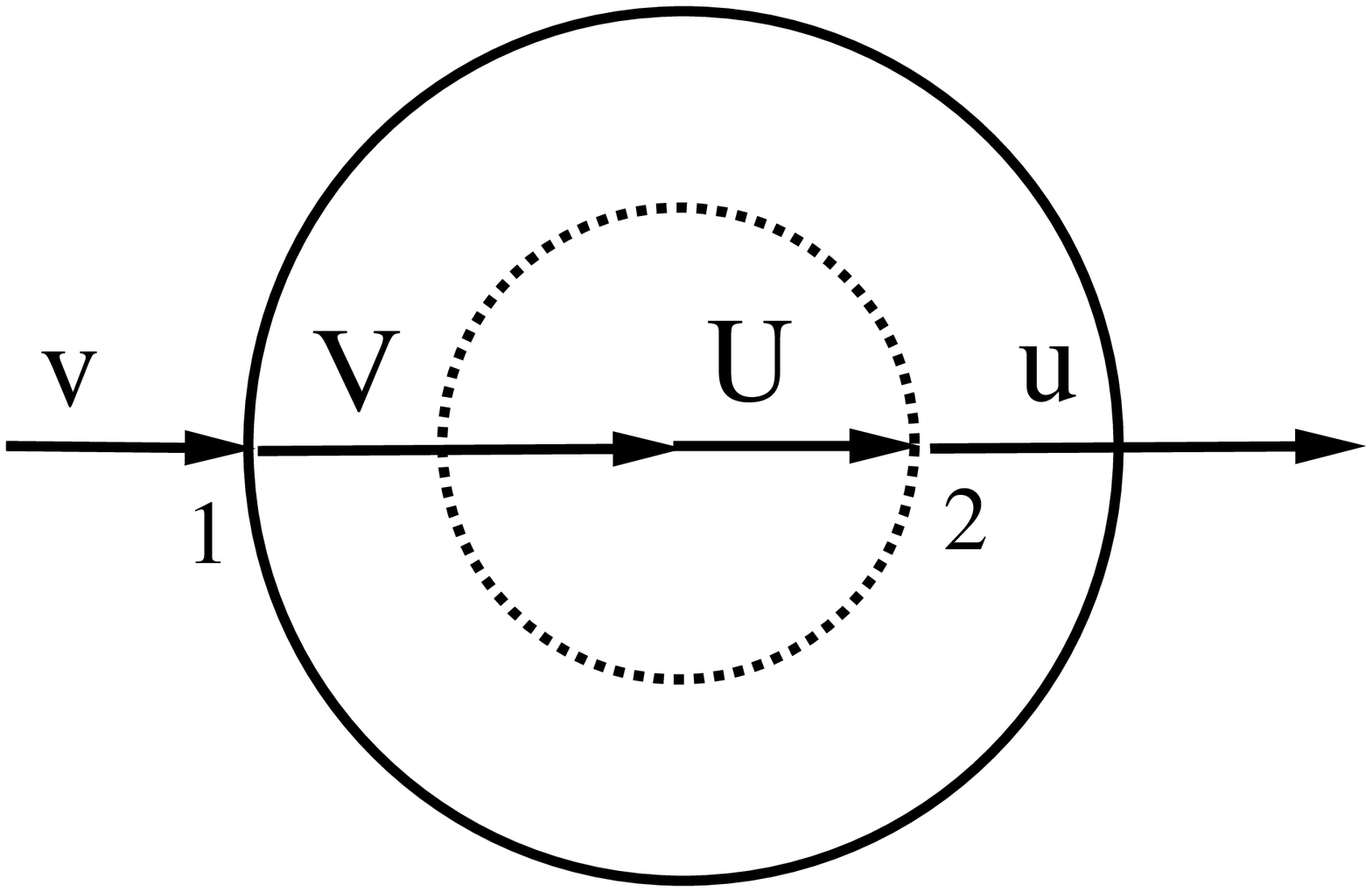}
\label{Figure 3}
\end{center}
\begin{caption}[]

An ingoing ray enters the collapsing shell at point $1$, passes through
the origin, and exits as an outgoing ray at point $2$, when the shell has
shrunk to a smaller radius (dotted circle).. This is illustrated schematically
in this diagram. Note that the rays in question are actually imploding
or exploding spherical shells of light.  
\end{caption}
\end{figure}

      There are now three conditions to be determined: the relation between
the values of the null coordinates $v$ and $V$ for the ingoing ray,
the relation between $V$ and $U$ at the center of the shell, and finally
the relation between $U$ and $u$ for the outgoing ray. This sequence of 
matchings is illustrated in Fig. 3.

\begin{itemize}
\item Let us suppose that our null ray enters the shell at a radius of
$R_1$, which is finitely larger than $2M$. At this point, both
$R/{R -2M}$ and  ${{dR}/ {dT}}$ are
finite and approximately constant. Thus   ${{dt}/ {dT}}$ is approximately
constant, so $t \propto T$. Similarly, $r^*$ is a linear function of $r$
in a neighborhood of $r=R_1$. Thus, we conclude that
\begin{equation}
V(v) =av+b   \label{eq:Vofv}
\end{equation}
in a neighborhood of $v=v_0$, where $a$ and $b$ are constants.
\item The matching of the null coordinates at the center of the shell is
very simple. Because $V=T+r$ and $U=T-r$, at $r=0$ we have that
\begin{equation}
U(V)=V.    \label{eq:UofV}
\end{equation}
\item We now consider the exit from the shell. We are interested in rays 
which exit when $R$ is close to $2M$. Let $T_0$ be the time at which
$R=2M$. (Note that this occurs at a finite time as seen by observers
{\it inside} the shell.) Then near $T=T_0$,
\begin{equation}
R(T) \approx 2M +A(T_0 -T),
\end{equation}
where $A$ is a constant. If we insert this into Eq. (\ref{eq:intrinsic}), 
we have that
\begin{equation}
\biggl({{dt}\over {dT}}\biggr)^2 \approx
\biggl({{R -2M}\over {2M}}\biggr)^{-2}\biggl({{dR}\over {dT}}\biggr)^2
\approx {{(2M)^2}\over {(T-T_0)^2}},
\end{equation}
which implies
\begin{equation}
t \sim -2M\, \ln\biggl({{T_0-T}\over B}\biggr), \qquad T\rightarrow T_0.
\end{equation}
Similarly, as $T\rightarrow T_0$, we have that
\begin{equation}
r^* \sim 2M\, \ln\biggl({{r -2M}\over {2M}}\biggr) \sim
         2M\, \ln\biggl[{{A(T_0 -T)}\over {2M}}\biggr],
\end{equation}
and hence that 
\begin{equation}
u = t - r^* \sim -4M\, \ln\Bigl({{T_0-T}\over B'}\Bigr).
\end{equation}
(Again, $B$ and $B'$ are constants.) However, in this limit we have that
\begin{equation}
U=T-r = T-R(T) \sim (1+A)T -2M -AT_0.
\end{equation}
\end{itemize}
\vspace{1cm}
Combining these results with Eqs. (\ref{eq:Vofv}) and (\ref{eq:UofV}) 
yields our final result, Eq. (\ref{eq:uofv}). Although we have performed
our explicit calculation for the special case of a thin shell, the result
is more general, as is revealed by Hawking's derivation. We can understand
why this is this case; the crucial logarithmic dependence which governs the
asymptotic form of $u(v)$ comes from the last step in the above sequence
of matchings. This step reflects the large redshift which the outgoing
rays experience after they have passed through the collapsing body, which
is essentially independent of the interior geometry. We could imagine dividing
a general spherically symmetric star into a sequence of collapsing shells.
As the null ray enters and exits each shell, each null coordinate is a linear
function of the preceeding one, until we come to the exit from the last shell.
At this point, the retarded time $u$ in the exterior spacetime is a
logarithmic function of the previous coordinate, and hence also a logarithmic
function of $v$, as given by Eq.~(\ref{eq:uofv}).
 
From Eq. (\ref{eq:asymp}), we see that the out-modes, when traced back 
to ${\cal I^{-}}$, have the form
\begin{equation}
F_{\omega\ell m} \sim \cases{e^{4Mi\omega\,\ln [(v_0 -v)/ C]},
                                            &$v < v_0$ \cr
                             0, &$v > v_0$. \cr}
\end{equation}
We can find the Bogolubov coefficients by Fourier transforming this function.
Recall that
\begin{equation}
F_{\omega\ell m} = \int_0^\infty d\omega' 
\Bigl( \alpha^*_{\omega' \omega\ell m} f_{\omega'\ell m} 
 - \beta_{\omega' \omega\ell m} f^*_{\omega'\ell m}\Bigr).
\end{equation}
Here we use the notation $\alpha_{\omega' \omega\ell m} =
\alpha_{\omega'\ell m, \omega\ell m}$ and $\beta_{\omega' \omega\ell m} =
\beta_{\omega'\ell -m, \omega\ell m}$, which is inspired by the fact 
that the dependence upon the angular coordinates must be the same for
each term in the above equation. Thus,
\begin{equation}
\alpha^*_{\omega' \omega\ell m} = {1\over{2\pi}}\sqrt{\omega'\over \omega}
\int_{-\infty}^{v_0} dv\, e^{i\omega' v}\,
e^{4Mi\omega\,\ln[(v_0 -v)/ C]},
\end{equation}
and 
\begin{equation}
\beta_{\omega' \omega\ell m} = -{1\over{2\pi}}\sqrt{\omega'\over \omega}
\int_{-\infty}^{v_0} dv\, e^{-i\omega' v}\,
e^{4Mi\omega\,\ln[(v_0 -v)/ C]},
\end{equation}
or, equivalently,
\begin{equation}
\alpha^*_{\omega' \omega\ell m} = {1\over{2\pi}}\sqrt{\omega'\over \omega}
e^{i\omega v_0}\int_0^\infty dv'\, e^{-i\omega' v'}\,
e^{4Mi\omega\,\ln(v'/C)},   \label{eq:alpha}
\end{equation}
and 
\begin{equation}
\beta_{\omega' \omega\ell m} = -{1\over{2\pi}}\sqrt{\omega'\over \omega}
e^{i\omega v_0}\int_0^\infty dv'\, e^{i\omega' v'}\,
e^{4Mi\omega\,\ln(v'/C)},   \label{eq:beta}
\end{equation}
where $v'=v_0 -v$.

Both of the above integrands are analytic everywhere except on the negative
real axis, where the branch cut of the logarithm function is located. Thus
\begin{equation}
\oint_C dv'\, e^{-i\omega' v'}\,e^{4Mi\omega\,\ln(v'/C)} =0, \label{eq:oint}
\end{equation}
where the integration is around the closed contour $C$ illustrated in
Fig. 4. We may now write
\begin{eqnarray}
\int_0^\infty dv'\, e^{-i\omega' v'}\,e^{4Mi\omega\,\ln(v'/C)}
&= &-\int_0^\infty dv'\, e^{i\omega' v'}\,e^{4Mi\omega\,\ln(-v'/C -i\epsilon)}
\nonumber \\
&= &-e^{4\pi M\omega}\int_0^\infty dv'\, 
               e^{i\omega' v'}\,e^{4Mi\omega\,\ln(v'/C)}. \label{eq:vints}
\end{eqnarray}
In the first step, we used Eq. (\ref{eq:oint}) and a 
$v' \rightarrow -v'$ change of
variables. In the second step, we used the relation $\ln(-v'/C -i\epsilon) =
-\pi i + \ln(v'/C)$. Comparison of this result with Eqs. (\ref{eq:alpha}) 
and (\ref{eq:beta}) leads to the result
\begin{equation}
|\alpha_{\omega' \omega\ell m}| = 
            e^{4\pi M\omega} |\beta_{\omega' \omega\ell m}|.
\end{equation}

\begin{figure}
\begin{center}
\leavevmode\epsfysize=6cm\epsffile{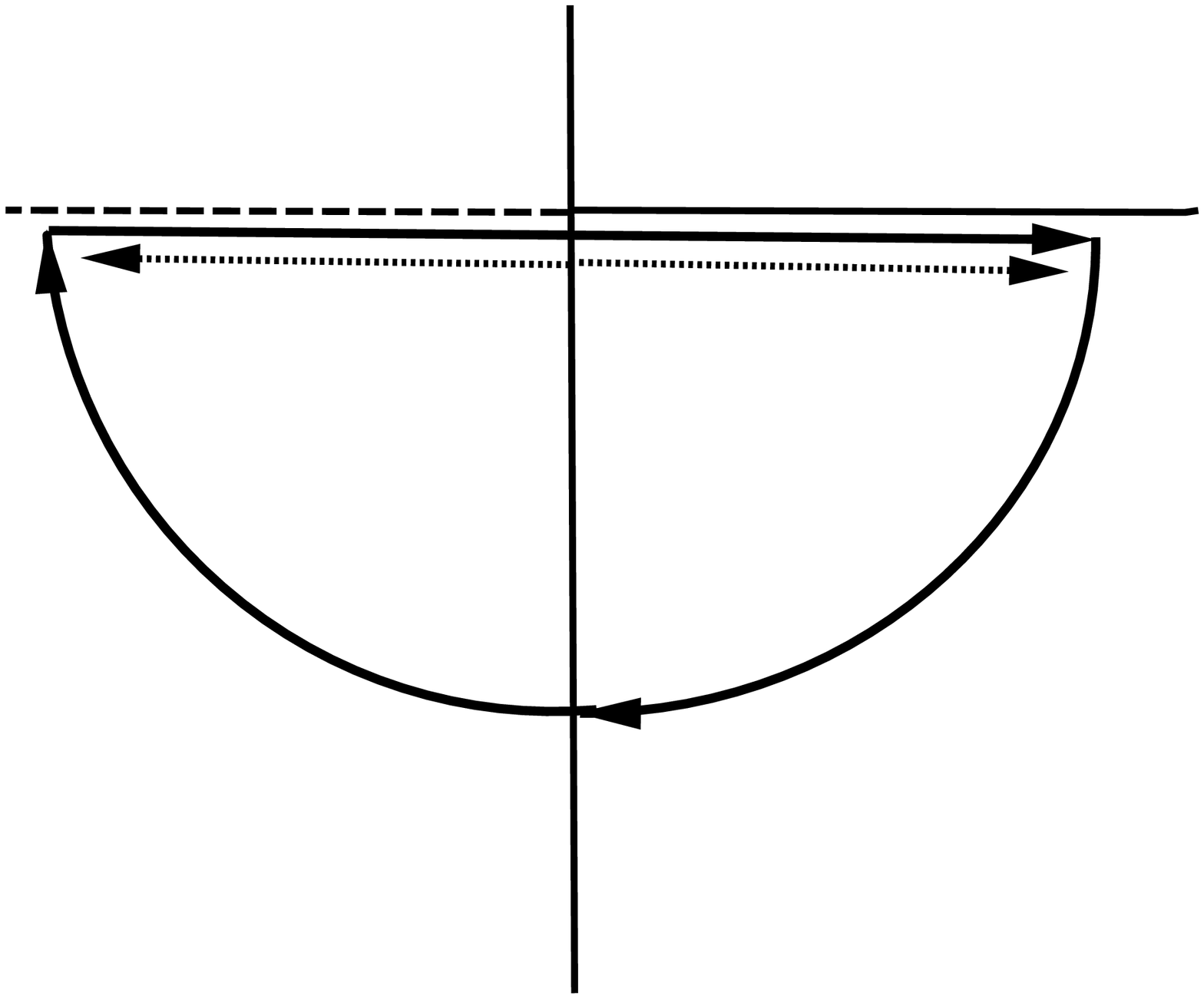}
\label{Figure 4}
\end{center}
\begin{caption}[]

The closed contour of the integration in Eq. (\ref{eq:oint})
is illustrated. The fact that this integral vanishes implies that the 
integrals along each of the dotted segments are equal, which implies
the first equality in Eq. (\ref{eq:vints}). 
\end{caption}
\end{figure}

The condition, Eq. (\ref{eq:alphabeta}), on the Bogolubov coefficients 
may be written as 
\begin{equation}
\sum_{\omega'} \Bigl(|\alpha_{\omega' \omega\ell m}|^2
                   - |\beta_{\omega' \omega\ell m}|^2 \Bigr) 
= \sum_{\omega'} \Bigl(e^{8\pi M\omega} - 1 \Bigr)
|\beta_{\omega' \omega\ell m}|^2  = 1.
\end{equation}
The mean number of particles created into mode $\omega\ell m$
is now given by
\begin{equation}
N_{\omega\ell m} = \sum_{\omega'}|\beta_{\omega' \omega\ell m}|^2 
                 = {1 \over{e^{8\pi M\omega} -1}}.
\end{equation}
This is a Planck spectrum with a temperature of
\begin{equation}
T_H = {1\over{8\pi M}},   \label{eq:Hawkingtemp}
\end{equation}
which is the Hawking temperature of the black hole. 

To show that these created particles produce a steady flow of energy to
${\cal I^{+}}$, we need to use either an analysis involving wavepackets
\cite{Hawking}, or else the following argument \cite{BD3}.
Note that the modes are discrete only if we regard the system
as being enclosed in a large box, which we may take to be a sphere of
radius ${\cal R}$. Then in the limit of large ${\cal R}$ we have
\begin{equation}
\sum_{\omega} \rightarrow {{\cal R}\over {2\pi}}\int_0^\infty d\omega.
\end{equation}
The total energy of the created particles is
 \begin{equation}
   E= \sum_{\omega \ell m} \omega N_{\omega\ell m} =
    {{\cal R}\over {2\pi}}\, \sum_{ \ell m} \,
      \int_0^\infty d\omega\,\omega N_{\omega\ell m}.
                                \label{eq:energy}
\end{equation}
This quantity would diverge in the limit that  ${\cal R} \rightarrow \infty$.
However, this simply reflects a constant rate of emission over an infinitely
long time (when backreaction of the radiation on the black hole is ignored).
We may find the luminosity by noting that it takes a time ${\cal R}$ for
an outgoing particle emitted by the black hole to reach the boundary of the 
spherical cavity. Thus, Eq. (\ref{eq:energy}) is the amount of energy 
emitted in a time ${\cal R}$, and the luminosity is
\begin{equation}
L = {E\over {\cal R}} =
{1\over {2\pi}}\, \sum_{ \ell m} \,
 \int_0^\infty d\omega\,\omega N_{\omega\ell m}.
\end{equation}
This result does not include the effect of the backscattering of the 
particles off of the spacetime curvature surrounding the black hole.
Let $\Gamma_{\ell m}$ denote the probability that a particle created
in mode $\ell m$ near the horizon escapes to infinity. Now our
expression for the luminosity becomes
\begin{equation}
L = {1\over {2\pi}}\, \sum_{ \ell m} \, \int_0^\infty d\omega\,\omega 
{{\Gamma_{\ell m}} \over {e^{8\pi M\omega} -1} }.
\end{equation}
The sum on $\ell$ converges because ${\Gamma_{\ell m}} \rightarrow 0$
as $\ell \rightarrow \infty$.

This result shows that black holes emit radiation with a (filtered)
Planckian spectrum. However, it may be shown that the radiation is 
indeed thermal \cite{Wald,Parker2,Hawking2}. This may be done, for example, 
by calculating the 
higher moments of the distribution, $\langle N_{\omega\ell m}^2\rangle$,
etc, and showing that they also take the forms required for thermal
radiation. This indicates that there are no correlations among the
emitted particles, at least in the semiclassical treatment given here. 
This thermal character of a black hole may be 
attributed to the loss of information across the event horizon. 

Thus one is lead to the subject of {\it black hole thermodynamics},
which was anticipated by Bekenstein \cite{Bekenstein} before Hawking's discovery.
With the Schwarzschild black hole temperature as given in 
Eq. (\ref{eq:Hawkingtemp}),
the First Law of Thermodynamics now takes the form
\begin{equation}
dS_{BH} = {dM \over {T_H}},
\end{equation}
where the black hole's entropy is given by $S_{BH} =4\pi M^2$ for the
Schwarzschild black hole, and more generally by ${1\over 4}A_H$, where
$A_H$ is the horizon area. The assignment of an entropy to the black
hole resolves the apparent paradox discovered by Bekenstein, who realized
that otherwise the Second Law of Thermodynamics would be violated when 
hot matter is thrown into a black hole. The Second Law now takes the
form
\begin{equation}
\Delta S = \Delta S_{BH} + \Delta S_{matter} \geq 0.
\end{equation}

Black hole evaporation provides a beautiful unification of aspects
of quantum theory, gravitation, and thermodynamics. However, there are still
a few unresolved issues. Among these are the questions of the use of 
ultra-high frequencies, and of the final state of black hole evaporation.
The ultra-high frequency issue arises when one calculates the frequency
$\omega'$ on ${\cal I^{-}}$ that is needed to produce the thermal radiation
at some time $t$ well after the collapse has occurred. 
From Eq. (\ref{eq:vofu}), this frequency is of the order of
\begin{equation}
\omega' = M^{-1} e^{t /{4M}}.
\end{equation}
If we take $t$ to be of the order of the expected black hole lifetime,
$M^3$, then $\omega' \approx M^{-1} e^{(M/M_{Pl})^2}$, where $M_{Pl}$
is the Planck mass. For a $1g$ black hole this leads to $\omega' \approx
{10}^{{10}^{10}} g$, which is enormously larger than the mass of the 
observable universe. A cutoff at any reasonable energy scale would appear
to quickly kill off the Hawking radiation. (See Ref. \cite{Jacobson} for a 
discussion of attempts to circumvent this conclusion.)

Finally there is the unresolved and much-debated issue of the final 
state \cite{Page}.
There seem to be three logical possibilities for the end result:
\begin{itemize}
\item {\bf A singularity.} This would mean that quantum theory does not
solve the classical singularity problem, and there is a loss of predictability.
Such an outcome would seem to imply that the theory is still incomplete.
\item {\bf A Stable Remnant.} If this remnant retained all of the information
which fell into the black hole during its history, it would appear to have
a huge number of internal degrees of freedom, and thus might give an
unacceptably large contribution to virtual processes. 
\item {\bf Total Evaporation.} This possibility seems to be the most
natural. It does lead to the conclusion that information is lost during
the black hole formation-evaporation process, unless there are subtle
correlations in the Hawking radiation that are not predicted by the
semiclassical theory presented in this lecture. Hawking has long taken the
viewpoint that there is both total evaporation and information loss 
\cite{Hawking76}. Other writers, including 't Hooft \cite{tHooft}, have argued
that interactions might imprint the needed correlations on the outgoing 
radiation so that the information slowly leaks out during the evaporation
process.
\end{itemize} 
It can safely be said  that this issue is not yet well understood at the 
present time. A more complete quantum theory of gravity will presumably be
needed to resolve these questions. One such candidate theory is string theory.
Progress has been made in calculating the entropy of near extreme black holes
in the context of string theory \cite{SV96}. 

Even in the absence of a complete quantum theory of gravity, one can attempt
to address such questions as the effects of quantum metric fluctuations
(The topic of Lecture 5.) on black hole evaporation. Some authors have made
estimates for the magnitude of the horizon fluctuations which, if correct,
would appear to invalidate the semiclassical derivation given earlier in this
Lecture \cite{Sorkin,Casher}. Other authors have argued for much smaller
fluctuations which would be consistent with the semiclassical treatment for
black holes whose mass is well above the Planck scale \cite{York,FS97}.
Again, this is an unresolved question. 

\vspace{1cm}

\section{GREEN'S FUNCTIONS AND $\langle T_{\mu\nu} \rangle$ \hfil\break
          IN CURVED SPACETIME}
\setcounter{equation}{0}
\renewcommand{\theequation}{3.\arabic{equation}}

  In this  lecture, we will discuss the removal of the ultraviolet 
divergences and the calculation of $\langle T_{\mu\nu} \rangle$ on
a curved background spacetime. In addition, we will discuss some
issues related to the infrared behavior of Green's functions in
curved spacetime. 

\subsection{Ultraviolet Behavior}

The ultraviolet divergences of a quantum field theory are related to the
short distance behavior of the vacuum expectation values of products of
field operators. We will be considering only free fields, so our primary
interest is in the two-point function. If one calculates a two-point 
function for some choice of the vacuum state in a curved spacetime, the
typical short distance behavior which one finds is
\begin{equation}
G(x,x') \sim {1 \over {2\pi^2}}\Bigl( -{1\over \sigma} + \ln\,\sigma
              + finite\,\,parts \Bigr), \qquad x'\rightarrow x,
\end{equation}
where $\sigma$ is one-half of the square of the geodesic distance 
between $x$ and $x'$. Thus, in flat spacetime or in the $x'\rightarrow x$
limit in curved spacetime, $\sigma = {1\over 2} (x-x')^2$. To be more
precise, let us discuss the {\it Hadamard function} for the scalar field
$\phi$:
\begin{equation}
G_1 (x,x') \equiv \langle 0|\phi(x) \phi(x')|0\rangle,
\end{equation}
where $|0\rangle$ is a chosen vacuum state. This function is said to 
have the {\it Hadamard form} if it can be expressed as
\begin{equation}
G_1(x,x') = {{U(x,x')}\over \sigma} + V(x,x')\: \ln\,\sigma +W(x,x'),
\end{equation}
where $U$, $V$, and $W$ are regular functions for all choices of $x$ and
$x'$. The functions $U$ and $V$ are geometrical quantities independent of the 
quantum state, and only $W$ carries information about the state. 
For most situations of interest to us, this form will hold for all
quantum states, and hence the singular part of $G_1(x,x')$ in the
coincidence limit will be state-independent. (The significance of the
Hadamard function {\it not} having the Hadamard form will be discussed below
when we deal with infrared divergences.)

We can formally construct the expectation value of the stress tensor,
$\langle T_{\mu\nu} \rangle$, as a limit of derivatives of $G^{(1)}$.
For example, consider the massless, minimally coupled scalar field, for
which
\begin{equation}
T_{\mu\nu} = \phi_{,\mu}\phi_{,\nu} -
               {1\over 2}g_{\mu\nu}\phi_{,\alpha}\phi^{,\alpha}.
\end{equation}
The formal expectation value of $T_{\mu\nu}$ is
\begin{equation}
\langle T_{\mu\nu} \rangle = {1\over 2} \lim_{x'\to x} \Bigl\{[
\partial_{\mu}\partial_{\nu'} -
 {1\over 2}g_{\mu\nu}\partial_{\alpha}\partial^{\alpha'} ]
              G^{(1)}(x,x') \Bigr\},  \label{eq:Tmunu}
\end{equation}
where $\partial_{\mu}$ denotes a derivative with respect to $x^\mu$
and $\partial_{\nu'}$ denotes one with respect to $x'^\nu$.
So far, this expression is only formal because it diverges in the
coincidence limit. However, for $x' \not= x$, it is a regularized
form of $\langle T_{\mu\nu} \rangle$. Here we are utilizing point
separation regularization \cite{DFCB}. There are several other methods for 
formally removing the ultraviolet divergences, including dimensional 
regularization and the zeta function method \cite{DC,Hawking3}. 
Point separation has the advantage of being
more generally applicable than do these other methods. Here we give a
brief summary of the basic ideas. For more details see, for example,
the books by Birrell and Davies \cite{BD} and by Fulling \cite{Fulling}. 
References which employ the properties of the Hadamard form include
Adler, et al \cite{ALN}, Wald \cite{Wald3}, Castagnino and Harari \cite{CH},
and  Brown and Ottewill \cite{BO}.
 
The right-hand side of Eq. (\ref{eq:Tmunu}) as it stands depends 
upon the direction
of the separation vector of the points $x$ and $x'$. This is undesirable,
and can be removed by averaging over these directions \cite{ALN}. If we do this, 
the asymptotic form for our regularized expression becomes
\begin{equation}
\langle T_{\mu\nu} \rangle \sim A\:{{g_{\mu\nu}}\over \sigma^2} + 
B\:{{G_{\mu\nu}}\over \sigma} + \bigl(C_1 H^{(1)}_{\mu\nu} +
C_2 H^{(2)}_{\mu\nu} \bigr)\: \ln\,\sigma.
\end{equation}
Here $A$, $B$, $C_1$, and  $C_2$ are constants, $G_{\mu\nu}$ is the Einstein 
tensor, and the $H^{(1)}_{\mu\nu}$ and $H^{(2)}_{\mu\nu}$ tensors are 
covariantly conserved tensors which are quadratic in the Riemann tensor.
Specifically, they are the functional derivatives with respect to the metric
tensor of the square of the scalar curvature and of the Ricci tensor, 
respectively:
\begin{eqnarray}
H^{(1)}_{\mu\nu} &\equiv& {1\over \sqrt{-g}} {\delta \over {\delta g^{\mu\nu}}} 
                                 \bigl[\sqrt{-g} R^2 \bigr] \nonumber \\
&=& 2\nabla_\nu \nabla_\mu R -2g_{\mu\nu}\nabla_\rho \nabla^\rho R
 - {1\over 2}g_{\mu\nu} R^2 +2R R_{\mu\nu},  \label{eq:H1}
\end{eqnarray}
and
\begin{eqnarray}
H^{(2)}_{\mu\nu} &\equiv& 
                {1\over \sqrt{-g}} {\delta \over {\delta g^{\mu\nu}}} 
            \bigl[\sqrt{-g} R_{\alpha\beta}R^{\alpha\beta} \bigr] 
= 2\nabla_\alpha \nabla_\nu R_\mu^\alpha - \nabla_\rho \nabla^\rho R_{\mu\nu}
\nonumber \\ &{}& -{1\over 2}g_{\mu\nu}\nabla_\rho \nabla^\rho R
  -{1\over 2}g_{\mu\nu} R_{\alpha\beta}R^{\alpha\beta} 
   +2R_\mu^\rho R_{\rho\nu}.    \label{eq:H2}
\end{eqnarray}

The divergent parts of $\langle T_{\mu\nu} \rangle$ may be absorbed by
renormalization of counterterms in the gravitational action. Write this
action as
\begin{equation}
S_G = {1\over {16\pi G_0}} \int d^4x\,\sqrt{-g}\, \Bigl( R -2\Lambda_0
      + \alpha_0 R^2 + \beta_0 R_{\alpha\beta}R^{\alpha\beta} \Bigr).
\end{equation}
We now include a matter action, $S_M$, and vary the total action, 
$S= S_G + S_M$, with respect to the metric. If we replace the classical
stress tensor in the resulting equation by the quantum expectation value,
$\langle T_{\mu\nu} \rangle$, we obtain the semiclassical Einstein equation
including the quadratic counterterms:
\begin{equation}
G_{\mu\nu} +\Lambda_0 g_{\mu\nu} +\alpha_0 H^{(1)}_{\mu\nu}
 +\beta_0 H^{(2)}_{\mu\nu} = -8\pi G_0 \langle T_{\mu\nu} \rangle.
\end{equation}
We may remove the divergent parts of $\langle T_{\mu\nu} \rangle$ in 
redefinitions of the coupling constants $G_0$, $\Lambda_0$, $\alpha_0$, 
and $\beta_0$. The renormalized values of these constants are then the physical
parameters in the gravitational theory. After renormalization, $G_0$ is 
replaced by $G$, the renormalized Newton's constant, which is the value actually
measured by the Cavendish experiment. Similarly, $\Lambda_0$ becomes the
cosmological constant $\Lambda$, which might be taken to be zero if we do 
not wish to have a cosmological term in the Einstein equations. 

In any case, the renormalized value of $\langle T_{\mu\nu} \rangle$ is 
obtained by subtracting the  terms which are divergent in the coincidence
limit. However, we are free to perform additional finite renormalizations
of the same form. Thus, $\langle T_{\mu\nu} \rangle_{ren}$ is defined only
up to the addition of multiples of the four covariantly conserved, geometrical
tensors $g_{\mu\nu}$, $G_{\mu\nu}$, $H^{(1)}_{\mu\nu}$, and $H^{(2)}_{\mu\nu}$.
Apart from this ambiguity, Wald \cite{Wald2} has shown under 
very general assumptions
that $\langle T_{\mu\nu} \rangle_{ren}$ is unique. Hence, at the end of the
calculation, the answer is independent of the details of the regularization 
and renormalization procedures employed.

An interesting feature of the renormalization of a quantum stress tensor is 
that it breaks conformal invariance. This leads to the conformal 
anomaly \cite{CD,DDI}.
A conformally invariant classical theory, such as electromagnetism or 
the conformally
coupled massless scalar field has the property that the trace of the stress
tensor vanishes: $T^\mu_\mu =0$. However, this classical property is lost 
in the renormalized quantum theory, and the expectation value of $T^\mu_\mu$
acquires a nonzero trace. This anomalous trace is independent of the choice 
of quantum state and is a local geometrical quantity. Furthermore, it is
not of a form which could be removed by a finite renormalization of the form
discussed above. For the case of the conformal ($\xi =1/6$) scalar field,
it is
\begin{equation}
\langle T^\mu_\mu \rangle_{ren} = -{1\over {2880\pi^2}} \Bigl(
R_{\alpha\beta\rho\sigma}R^{\alpha\beta\rho\sigma} 
- R_{\alpha\beta}R^{\alpha\beta} - \nabla_\rho \nabla^\rho R \Bigr).
                                            \label{eq:trace}
\end{equation}

Although the conformal anomaly is a state independent object, in general
$\langle T_{\mu\nu} \rangle_{ren}$ is a state-dependent quantity. This is,
of course, necessary so that it carry information about the matter content 
of particular quantum states. Because of this state-dependence, it is not
possible to make any general statements concerning its order of magnitude.
However, one typically finds for states which have the appearance  of a 
``vacuum state'' (i.e. in some sense, states of minimum excitation) that
the order of magnitude (in a local orthonormal frame) is 
\begin{equation}
\langle T_{\mu\nu} \rangle_{ren} \approx C \ell^{-4},
\end{equation}
where $\ell$ is the characteristic local radius of curvature of the 
spacetime, and $C$ is a dimensionless constant which tends to be of the
order of $10^{-3}$ to $10^{-4}$. For example, in the Einstein universe
the energy density for a massless conformal scalar field is \cite{F75}
\begin{equation}
\langle T_{tt} \rangle_{ren} = {1\over {480\pi^2 a^4}}
\end{equation}
in the vacuum state, which in this case is the unique state of lowest energy
due to the presence of a global time-like Killing vector. This case is
also of interest because the conformal anomaly, Eq. (\ref{eq:trace}), 
vanishes here.

\subsection{Infrared Behavior}

In our discussion of the Hadamard form, we noted that it is a common, although
not universal property of quantum states. In a state in which the two-point
function does not have the Hadamard form, the renormalization procedure 
outlined above will not remove all of the infinities from the stress tensor.
In flat spacetime, a state which does not have the Hadamard form would have
to be considered to be unphysical if the normal-ordered energy density were
infinite. Fulling, Sweeny and Wald \cite{FSW} have shown that a 
two point function which has the
Hadamard form at one time will have it at all times. In particular, in any
spacetime which is asymptotically flat in the past or in the future, the 
Hadamard form will hold if it holds in the flat region. Thus, it seems 
reasonable to require that the two point function having the Hadamard form 
be a criterion for a physically acceptable state.

Examples of states which do not have the Hadamard form may be constructed
even in flat spacetime \cite{FP}. Let us first consider a massless scalar field 
in flat four-dimensional spacetime, which has the mode expansion
\begin{equation}
\varphi = \sum_{\bf k} (a_{\bf k} f_{\bf k} + a^\dagger_{\bf k} f^*_{\bf k}),
\end{equation}
where we now take our modes to have the form (box normalization in a volume
$V$)
\begin{equation}
f_{\bf k} ={{e^{i{\bf k\cdot x}}} \over 
{\sqrt {2\omega V}}}\, \bigl[\alpha(\omega)e^{-i\omega t}
                            +\beta(\omega)e^{i\omega t} \bigr].
\end{equation}
In order that $f_{\bf k}$ have unit norm, we must require that
\begin{equation}
|\alpha(\omega)|^2 - |\beta(\omega)|^2 =1. \label{eq:norm}
\end{equation}
This expansion defines a state $|\psi\rangle$ such that 
$a_{\bf k} |\psi\rangle=0$. This state is the vacuum state only if 
$\beta =0$. Nonetheless, we may still define a two-point function by
\begin{eqnarray}
\langle \psi|\phi(x) \phi(x')|\psi\rangle &=& {1\over {2(2\pi)^3}}
\int d^3k \,\omega^{-1}\, \Bigl\{ \bigl[\alpha(\omega)e^{-i\omega t}
               +\beta(\omega)e^{i\omega t} \bigr] \nonumber \\
&{}&  \times \bigl[\alpha^*(\omega)e^{i\omega t'}
+\beta^*(\omega)e^{-i\omega t'} \bigr] e^{i{\bf k}\cdot ({\bf x}-{\bf x'})}
\Bigr\}\, .
\end{eqnarray}
Let us suppose that the integral is dominated by low frequency modes. Then
\begin{equation}
\langle \psi|\phi(x) \phi(x')|\psi\rangle \sim  {1\over {(2\pi)^2}}
\int d\omega\,\omega |\alpha(\omega) + \beta(\omega)|^2. \label{eq:lowfreq}
\end{equation}
There exist choices of the functions $\alpha(\omega)$ and $\beta(\omega)$
which satisfy Eq. (\ref{eq:norm}), but for which this integrand diverges as 
$\omega \rightarrow 0$. For example, let
\begin{equation}
\beta(\omega)=\omega^{-c}, \qquad \alpha(\omega)= (1+\omega^{-2c})^{1\over 2}.
\end{equation}
In this case
\begin{equation}
|\alpha(\omega) + \beta(\omega)| \sim \omega^{-c}, \qquad \omega \rightarrow 0,
\end{equation}
and the two-point function is infinite for all
$x$ and $x'$ if $c>1$. This is an example of an infrared divergence.
The result that the Hadamard form is preserved shows that infrared divergences
will not arise during the course of time-evolution from a state which is 
free of them. Thus we are justified in excluding such states as unphysical.

In the above example, it may seem that we had to go to some lengths to find
pathological states. However, in other spacetimes, the infrared divergences
appear in apparently natural choices of quantum state, 
and the cure is remarkably similar to the
prescription which caused the problems in the above example. Let us now
consider a massless scalar field in two-dimensional spacetime. If we follow
a construction parallel to that given above in four dimensions, we find
that the analog of Eq. (\ref{eq:lowfreq}) is now 
\begin{equation}
\langle \psi|\phi(x) \phi(x')|\psi\rangle \sim  {1\over {4\pi}}
\int d\omega\,\omega^{-1} |\alpha(\omega) + \beta(\omega)|^2.
\end{equation}
If we were to choose the Minkowski vacuum state, for which $\alpha =1$
and $\beta =0$, we have an infrared divergence. This is a well-known
property of massless fields in two dimensions. However, there exist states 
which are free of infrared divergences. For example, let
\begin{equation}
\beta(\omega)=-\omega^{-c}, \qquad \alpha(\omega)= 
   (1+\omega^{-2c})^{1\over 2}.    \label{eq:IRfree}
\end{equation}
Now $|\alpha(\omega) + \beta(\omega)| \sim {1\over 4}\omega^{2c}$ as
$\omega \rightarrow 0$, and the two-point function is finite if $c>0$.
Thus the infrared divergences in two dimensions are the consequence of
a poor choice of quantum state and are remedied when a better choice is made.
Note that the physically allowable quantum states are all ones which break 
Lorentz invariance. One may show that in any state which is free of infrared
divergences, $\langle \phi^2\rangle$ must be a growing function of 
time \cite{FV86}. In particular, in the quantum state defined by 
Eq. (\ref{eq:IRfree}), one finds
\begin{equation}
\langle \phi^2\rangle \sim t^{2c}, \qquad t\rightarrow \infty.
\end{equation}

Another example in which similar behavior occurs is deSitter spacetime.
In the representation as a spatially flat Robertson-Walker universe, its
metric is
\begin{equation}
ds^2 = {1\over{(H\eta)^2}} \bigl(d\eta^2 -d{\bf x}^2 \bigr)
     = dt^2 - e^{2Ht} d{\bf x}^2.
\end{equation}
These coordinates cover one-half of the full deSitter space, but this
is not a serious problem for our purposes. In the context of inflationary
models, one is interested in spacetimes which involve only a piece of 
deSitter space. The massless, minimally coupled scalar field, which satisfies
the wave equation
\begin{equation}
\Box \phi =0,
\end{equation}
has solutions in terms of Hankel functions \cite{Bunch&Davies}:
\begin{equation}
f_{\bf k} \propto e^{i{\bf k\cdot x}} \Bigl[ c_2 H^{(2)}_{3\over 2}(k\eta)
                                 + c_1 H^{(1)}_{3\over 2}(k\eta) \Bigr].
\end{equation}
where $|c_2|^2 - |c_1|^2 =1$. The vacuum state which is invariant under 
the action of the deSitter symmetry group is given by
\begin{equation}
c_2 =1, \qquad c_1 =0.
\end{equation}
However, because $H^{(2)}_{3\over 2}(k\eta) \sim k^{- {3\over 2}}$ as
$k \rightarrow 0$, this state is infrared divergent. As before, we may
find states which are free of such divergences; here what is required is
a choice of $c_1(k)$ and  $c_2(k)$ such that $|c_1(k) + c_2(k)| \rightarrow 0$
as $k \rightarrow 0$. Such states necessarily break deSitter invariance
and lead to growth \cite{FV82b,Linde,Staro} of $\langle \phi^2\rangle$:
\begin{equation}
\langle \phi^2\rangle \sim {{H^3 t}\over {4\pi^2}}, 
              \qquad t\rightarrow \infty.   \label{eq:growth}
\end{equation}
This is similar to the result in two-dimensional flat spacetime, although
now the asymptotic rate of growth is independent of the choice of state, 
so long as the state is well defined.

This growth of $\langle \phi^2\rangle$ in deSitter space has consequences
for inflationary models. For example, in the ``new inflation'' model,
one postulates a self-coupled scalar field $\phi$ with a potential $V(\phi)$
which is very flat near the origin. It is the long period of slow rolling
away from the origin which allows sufficient inflation to solve the horizon
problem. During this period, $\phi$ approximately satisfies 
Eq. (\ref{eq:growth}), so the root-mean-square value of $\phi$ must 
grow as $\sqrt{t}$. This tends to limit the period of inflation.

Another application of these results is to models of global symmetry breaking.
Let us consider the Goldstone model of $U(1)$ symmetry breaking, where a
complex scalar field $\Phi$ has the Lagrangian density
\begin{equation}
{\cal L} = \partial_{\alpha}\Phi^* \partial^{\alpha}\Phi - V(\Phi),
\end{equation}
where 
\begin{equation}
V(\Phi) = -{1\over 2} m^2 \Phi^*\Phi +
           {1\over 4} \lambda \bigl(\Phi^*\Phi \Bigr)^2.
\end{equation}
This potential has an unstable maximum at $\Phi =0$, but minima at
\begin{equation}
\Phi = \sigma e^{i\phi}, \qquad \sigma =m \lambda^{-1/2}.
\end{equation}
If $\sigma$ is constant, then the equation of motion for $\Phi$ implies
that $\Box \phi =0$. Thus the Goldstone boson $\phi$ is a massless
scalar field. 

We now wish to treat $\phi$ as a quantized field and calculate the 
expectation value of $\Phi$. This requires that we find the expectation value
of the exponential of an operator. Decompose $\phi$ into its positive and 
negative frequency parts: $\phi = \phi^{+} + \phi^{-}$, where 
$\phi^{+}|0\rangle =0$ and $\langle 0|\phi^{-} =0$. In terms of annihilation
and creation operators, $\phi^{+} = \sum_j a_j f_j$ and $\phi^{-} =
\sum_j a^\dagger_j f^*_j$. We now write
\begin{equation}
e^{i\phi} = e^{i(\phi^{+} + \phi^{-})} = 
e^{i\phi^{-}} e^{-{1\over 2}[\phi^{+}, \phi^{-}]} e^{i\phi^{+}},
\end{equation}
where in the second step we use the Campbell-Baker-Hausdorff formula.
We now take the vacuum expectation value of this expression and use the
facts that $e^{i\phi^{+}}|0\rangle = |0\rangle$ and $\langle 0|e^{i\phi^{-}}  
= \langle 0|$, which follow immediately if the exponentials are expanded in
a power series. Finally, we use $[\phi^{+}, \phi^{-}] =\sum_j f_j f^*_j
= \langle \phi^2 \rangle$ to write 
\begin{equation}
\Bigl\langle \Phi \Bigr\rangle = \sigma \Bigl\langle e^{i\phi} \Bigr\rangle
= \sigma e^{-{1\over 2}\langle \phi^2 \rangle}.  \label{eq:eiphi}
\end{equation}
The ultraviolet divergence in $\langle \phi^2 \rangle$ is understood to be 
absorbed in a rescaling of $\Phi$ (a wavefunction renormalization).
In spacetimes, such as four dimensional flat space, where one can have
$\langle \phi^2 \rangle$ constant in a physically acceptable state, then
there are stable broken symmetry states in which $\langle \Phi \rangle
\not= 0$. However, in two dimensional flat spacetime or in four dimensional
deSitter spacetime, the growth of $\langle \phi^2 \rangle$ forces
$\langle \Phi \rangle$ to decay in time: $\langle \Phi \rangle \rightarrow
0$ as $t \rightarrow \infty$. In these cases, the infrared behavior of the
massless scalar field prevents the existence of a stable state of broken
symmetry.

\vspace{1cm}

\section{NEGATIVE ENERGY DENSITIES \hfil\break AND FLUXES}
\label{sec:negen}
\setcounter{equation}{0}
\renewcommand{\theequation}{4.\arabic{equation}}

       This  lecture will discuss one of the special properties of the
local energy density in  quantum field theory, namely, that it can be 
negative. Negative energy is crucial for an understanding of the Hawking 
effect, in that a negative energy flux across the horizon is needed to 
implement the backreaction of the spacetime metric to the outgoing radiation.
(Note that one cannot think of the backreaction as simply due to the 
positive energy outgoing radiation, as such radiation would undergo an
infinite blueshift when traced back to $r=2M$.) Rather, one may think of
pairs of particles being created in the region outside $r=2M$, one member
of the pair escapes to infinity, and the other falls into the horizon.
The latter particle carries negative energy as measured at infinity.
This picture is consistent with calculations of the quantum field stress
tensor near the black hole horizon in both two \cite{DFU} and 
four \cite{Elster,JMO} dimensional models.
 
However, negative energy densities and fluxes arise even in flat spacetime.
A simple example is the Casimir effect \cite{Casimir}, 
where the vacuum state of the 
quantized electromagnetic field between a pair of conducting plates separated
by a distance $L$ is a state of constant negative energy density
\begin{equation}
\rho = \langle T_{tt} \rangle = -{{\pi^2}\over{720 L^4}}.
\end{equation}
Negative energy density can also arise as the result of quantum coherence
effects, which will be the principal concern of this lecture. Although
we will restrict our attention to free fields in Minkowski spacetime, the 
basic considerations are much more general. In fact, it may be shown under
very general assumptions that all quantum field theories admit states
for which the energy density may be arbitrarily negative at a given 
point \cite{EGJ,Kuo97}. We have already met one example of negative energy in the
moving mirror models discussed in Sec. \ref{sec:mirror}. From 
Eq.~(\ref{eq:mirror3}), we see that if the acceleration on the mirror is 
increasing to the right, then the flux of energy radiated to the right is 
negative.

We can illustrate the basic phenomenon of negative energy arising from
quantum coherence with a very simple example. Let 
the quantum state of the system be a superposition of the vacuum and a 
two particle state:
\begin{equation}
|\Psi\rangle = 
{1\over\sqrt{1+\epsilon^2}} (|0\rangle + \epsilon|2\rangle). 
                                              \label{eq:vacplus2}
\end{equation}
Here we take the relative amplitude $\epsilon$ to be a real number. Let
the energy density operator be normal-ordered:
\begin{equation}
\rho = :T_{tt}:\, ,
\end{equation}
so that $\langle0|\rho|0\rangle =0.$ Then the expectation value of the energy
density in the above state is
\begin{equation}
\langle \rho \rangle = {1 \over {1+\epsilon^2}} 
      \Bigl[2 \epsilon {\rm Re}(\langle0|\rho|2\rangle)
            + \epsilon^2 \langle2|\rho|2\rangle \Bigr]\, .
\end{equation}
We may always choose $\epsilon$ to be sufficiently small that the first
term on the right hand side dominates the second term. However, the former
term may be either positive or negative. At any given point, we could choose
the sign of $\epsilon$ so as to make $\langle \rho \rangle  < 0$ at that point.

Note that the integral of $\rho$ over all space is the Hamiltonian, which 
does have non-negative expectation values:
\begin{equation}
\langle H \rangle = \int d^3 x \langle \rho \rangle \geq 0.
\end{equation}
In the above {\it vacuum + two particle} example, the matrix element 
$\langle0|\rho|2\rangle$, which gives rise to the negative energy density, 
has an integral over all space which vanishes, so only $\langle2|\rho|2\rangle$
contributes to the Hamiltonian.

This example is a limiting case of a more general class of quantum states
which may exhibit negative energy densities, the squeezed states.
A general squeezed state for a single mode can 
be expressed as \cite{Caves,Mandel}
\begin{equation}
|z,\zeta\rangle=D(z)\,S(\zeta)
\,|0\rangle,
\end{equation}
where $D(z)$ is the displacement operator
\begin{equation}
D(z)\equiv \exp(z a^{\dagger}-
z^{\ast}a)=e^{-|z|^2/2}\,
e^{z a^\dagger}\,e^{-z^\ast a}
\end{equation}
and $S(\zeta)$ is the squeeze operator 
\begin{equation}
S(\zeta)\equiv \exp[{1\over 2}\zeta^\ast a^2
-{1\over 2}\zeta ({a^\dagger})^2].
\end{equation}
Here 
\begin{equation}
z = s e^{i\gamma}
\end{equation}
 and
\begin{equation}
\zeta = re^{i\delta}.
\end{equation}
are arbitrary complex numbers.
The displacement and squeeze operators satisfy the relations
\begin{equation}
D^{\dagger}(z)\,a\,D(z)=a+z,
\end{equation}
\begin{equation}
D^{\dagger}(z)\,a^{\dagger}\,D(z)
=a^{\dagger}+z^{\ast},
\end{equation}
\begin{equation}
S^{\dagger}(\zeta)\,a\,S(\zeta)=
a\,\cosh r-a^{\dagger}e^{i\delta}\sinh r, \label{eq:squeeze}
\end{equation}
and
\begin{equation} 
S^{\dagger}(\zeta)\,a^{\dagger}\,S(\zeta)=
a^{\dagger}\,\cosh r-ae^{-i\delta}\sinh r.
\end{equation}

When $\zeta=0$ we have the familiar coherent states, $|z\rangle=
|z,0 \rangle$, which describe classical excitations. In such a 
state, the expectation value of a quantum field $\phi$ is
\begin{equation}
\langle \phi \rangle =   z f + z^* f^*,
\end{equation}
where $f$ is the mode function for the excited mode. This is a solution
of the classical wave equation. Furthermore, the quantum fluctuations in this
state are minimized:
\begin{equation}
\langle :\phi^2: \rangle = \langle \phi \rangle^2.
\end{equation}
 
The opposite limit from a coherent state is a ``squeezed vacuum state'',
$|0,\zeta\rangle$, for which $z=0$. Sufficiently squeezed states can
exhibit negative energy density, and a squeezed vacuum state always
has $ \langle \rho \rangle < 0$ somewhere. 
One may think of the effect of the squeezing as decreasing  the quantum 
uncertainty in one variable, but increasing it in the conjugate variable.
Squeezed vacuum states are
of particular interest to us because they are the states which arise as 
a result of quantum particle creation. That is, the in-vacuum state is
a squeezed vacuum state in the out-Fock space.  

We may illustrate this by considering a Bogolubov transformation involving
a single mode. Let
\begin{equation}
a = \alpha^* b - \beta^* b^\dagger,
\end{equation}
or, equivalently,
\begin{equation}
b = \alpha a + \beta^* a^\dagger,
\end{equation}
where $|\alpha|^2 -|\beta|^2 =1$. These are single mode versions of
Eqs. (\ref{eq:Bogo1}) and (\ref{eq:Bogo2}). 
As before, the in-vacuum satisfies $a|0\rangle_{in}=0$,
and the out-vacuum satisfies $b|0\rangle_{out}=0$. We wish to express
$|0\rangle_{in}$ as a state in the out-Fock space. This may achieved
by the action of some operator $\Sigma$ upon $|0\rangle_{out}$:
\begin{equation}
|0\rangle_{in} = \Sigma |0\rangle_{out}.
\end{equation}
Act with the operator $\Sigma^\dagger a$ on both sides of this equation
to obtain
\begin{equation}
\Sigma^\dagger a \Sigma \,|0\rangle_{out} =0.
\end{equation}
Hence we may identify $\Sigma^\dagger a \Sigma = b =
\alpha a + \beta^* a^\dagger.$ However, this is basically of the same 
form as Eq. (\ref{eq:squeeze}). We may choose the phase of our mode 
so that $\alpha$
is real. Then, if we let $r$ and $\delta$ be such that $\alpha = \cosh\, r$,
and $\beta = -e^{-i\delta} \sinh\, r$, we see that $\Sigma = S$, the
squeeze operator. Hence $|0\rangle_{in}$ is a squeezed vacuum state in
the out-Fock space.

Squeezed states of light have recently been created in the laboratory
by use of nonlinear optics \cite{K}. 
The essential idea is that a nonlinear medium
acts like a material with a time-dependent dielectric function when a 
strong, time-varying classical electromagnetic field is applied. Photon
modes propagating through a time-dependent dielectric will undergo a
mixing of positive and negative frequencies, just as in a time-dependent
gravitational field, and photons will be quantum mechanically created into
a squeezed vacuum state. Let us examine how this happens in more detail.
In a nonlinear material, the magnitudes of the displacement $D$ and of the
electric field $E$ cease to be proportional to one another and become
nonlinearly related. Typically the departure from linearity is small, and
may be approximated by a quadratic function:
\begin{equation}
D \approx \epsilon_0 E + \epsilon_1 E^2 + \cdots \,.
\end{equation}
We now write the electric field as
\begin{equation}
E = E_0 + E_1 \,,
\end{equation}
where $E_0 = E_0({\bf x},t)$ is the strong classical field, e.g. an intense
laser field, and $E_1$ is a test field with $|E_1| \ll |E_0|$. To first
order in $E_1$, we find that
\begin{equation}
D_1 = D - D_0 \approx 
               \bigl[\epsilon_0  + \epsilon_1 E_0({\bf x},t) \bigr] E_1 \,.
\end{equation}
Thus the test field behaves to this order as though it were propagating in
a linear material with a time dependent dielectric function:
\begin{equation}
\epsilon_{eff}({\bf x},t) = \epsilon_0  + \epsilon_1 E_0({\bf x},t) \,.
\end{equation}
Thus a pure positive frequency mode which enters the material  
will typically exit it as a superposition of positive and negative
frequency parts. Consequently, photons are created into a squeezed vacuum
state, just as they might be in a time dependent gravitational field.
 Thus, the observation of squeezed states of photons
can be regarded as an experimental confirmation of the formalism developed
in Lecture I. 

In the present context, squeezed states are of particular interest as 
examples of
negative energy density states. Although it has not been possible to
directly detect negative energy densities in the laboratory, the reduction
of quantum noise due to squeezing has been observed \cite{K}. We can think of 
negative energy density as a related reduction. When the quantum fluctuations 
are momentarily suppressed so that the energy density falls below the vacuum
level, we have negative energy density. This suppression of quantum
fluctuations could in principle be detected by a spin system, and would
manifest itself in the net magnetic moment of the system increasing
{\it above} the vacuum value \cite{FGO}. We can visualize this as occurring
because the vacuum fluctuations tend to depolarize the spins, and negative
energy tends to reduce this depolarizing effect, allowing the spins to
become more perfectly aligned. Grove \cite{Grove} has shown that in some
cases one may use a switched particle detector as a model negative energy
detector. The act of switching the detector on and off by itself can
create an excitation of the detector. The effect of the negative energy
is to suppress the excitations that would otherwise occur. 
 
A closely related phenomenon to a negative energy density is a negative 
energy flux. This arises when we have a quantum state in which all of the
particles are moving in one direction, but the instantaneous flow of
energy is in the opposite direction. Such a state would also have a locally
negative energy density, but not all states with negative energy density
carry a negative energy flux. The examples given above for negative energy 
density also exhibit a negative energy flux if the mode in question is a
travelling wave mode (as opposed to a standing wave). Another example of
a negative energy flux arises in moving mirror models in two-dimensional
spacetime \cite{FD}. Here if the acceleration of the mirror is increasing in the 
direction of the observer, then the mirror emits a negative energy flux.

      If one could have arbitrary fluxes of negative energy, 
it would seem that one could shine the negative energy on a hot object
and cause a net decrease in entropy, and hence violate the second law of
thermodynamics \cite{F78}. For example, the object could be a 
black hole. The absorption of negative energy would seem to decrease
the black hole's entropy without a compensating increase in the entropy
of radiation. 

However, there are some constraints upon negative energy fluxes. The first
is that the net energy must be non-negative. Let $F(t)$ be the instantaneous
flux. Then
\begin{equation}
\int_{-\infty}^{\infty} F(t)\,dt \, \geq 0.
\end{equation}
This inequality alone is not sufficient to prevent an arbitrarily large
violation of the second law from occurring before the compensating positive
energy arrives. There are stronger restrictions on negative energy fluxes
which constrain the magnitude and duration of a pulse of negative energy.
In flat two-dimensional spacetime such an inequality is of the form
\begin{equation}
|F| < {t^{-2}},  
\end{equation}
where $|F|$ is
the magnitude of the negative flux and $t$ is its duration. This inequality 
implies that $|F|t$, the amount of negative energy which passes by a fixed 
location in time 
$t$ is less than the quantum energy uncertainty on that timescale, $t^{-1}$.  
A more precise version of this inequality is obtained by multiplying $F$ by 
a peaked function of time
whose time integral is unity and whose characteristic width is $t_0$. A suitable
choice of such a function is $t_0/[\pi(t^2+{t_0}^2)]$. The inequality 
is \cite{F91}
\begin{equation}
 {\hat F} \equiv {t_0 \over \pi} \int_{-\infty}^{\infty} 
{{F(t) dt}\over {t^2+{t_0}^2}}  \geq -{1 \over {16 \pi {t_0}^2}}\, .
                                           \label{eq:flux}
\end{equation}
which holds for any quantum state for which only modes with $k>0$ are
excited. (This restriction is needed in order to distinguish a flow of
negative energy to the right from a flow of positive energy to the left.)
An illustration of the application of this inequality is afforded by the case
of a negative energy pulse followed at a later time by a compensating 
positive energy pulse.
The most efficient separation of
positive and negative energy is obtained by delta-function pulses. Consider
the following flux:
\begin{equation}
F(t) = |\Delta E| \bigl[ -\delta(t) + \delta(t-T) \bigr].  
\end{equation}
This represents a pulse of negative energy followed a time $T$ later by an 
exactly compensating pulse of positive energy. The inequality, 
Eq. (\ref{eq:flux}), yields
\begin{equation}
|\Delta E| \leq {{{T^2 +{t_0}^2} \over 16{t_0}T^2}}\,. \label{eq:fluxQI}  
\end{equation}
This relation is true for all $t_0$, but the best constraint on $|\Delta E|$
is obtained by setting $t_0 = T$. Then we find
\begin{equation}
|\Delta E| \leq {1 \over {8T}}.    
\end{equation}
This inequality tells us that there is a maximum separation in time between
the two pulses, which is within the limits allowed by the uncertainty principle.

   Recall that one mechanism for producing a flux of negative energy is a
moving mirror, as discussed in Sec. \ref{sec:mirror}. From 
Eq.~(\ref{eq:mirror2}) we can see why there is an inverse relation between the
magnitude and duration of the flux in this case. The energy radiated to the right
is negative only when the acceleration to the right is increasing. However, this 
means that an inertial observer located to the right of the mirror will 
eventually collide with the mirror, unless this situation changes, and the
mirror starts to accelerate to the left. Furthermore, the larger is $\dot a$, 
the sooner this collision will occur. One may show that Eq.~(\ref{eq:fluxQI})
is satisfied in this case. (See Fig. 7 of Ref. \cite{FR90}.)

An inequality similar to Eq.~(\ref{eq:flux}) applies to the massless scalar 
field in four-dimensional flat spacetime:
\begin{equation}
 {\hat F_x} \equiv {t_0 \over \pi} \int_{-\infty}^{\infty} 
{{{F_x}(t) dt}\over {t^2+{t_0}^2}} \geq -{3 \over {32{\pi^2} 
{t_0}^4 }}.  \label{eq:4dflux}
\end{equation}
Here ${F_x}(t) = \langle T^{xt} \rangle$ is the flux in the x-direction,
and the expectation value is taken in any quantum state in which only modes 
with $k_x \geq 0$ are excited. 
Again we can apply it to the case of separated negative and positive energy
pulses. Let
\begin{equation}
F_x(t) = {|\Delta E|\over A}\bigl[ -\delta(t) + \delta(t-T) \bigr].
                                         \label{eq:deltaflux}
\end{equation}
This represents a plane delta function pulse of negative energy which has 
a magnitude  $|\Delta E|$ over a collecting area $A$, and which is followed
a time $T$ later by compensating positive energy. Here we may regard $T$ as 
being the timescale for the duration of the negative energy. In order that 
the all parts of the collecting system be causally connected on a time $T$, 
this time should be larger than the linear dimensions of the collector.
If we insert Eq. (\ref{eq:4dflux}) into Eq. (\ref{eq:deltaflux}), 
set $t_0 = T$ and require that $A\leq T^2$, then we find that 
\begin{equation}
|\Delta E| \leq {3 \over {16\pi T}}.    \label{eq:Qineq}
\end{equation}
Again there is a constraint which requires the positive energy to arrive 
within a time $1/|\Delta E|$.

Similar inequalities may be proven in black hole spacetimes \cite{FR90,FR92}. 
Let us consider 
an attempt to create a naked singularity using negative energy. We could
start with an extreme, $Q=M$, charged black hole. We then shine some negative 
energy on it so as to decrease $M$ with no change in the charge $Q$. This should
result in the naked singularity Reissner-Nordstr{\o}m spacetime. 
Even if subsequent
positive energy converts it back into a black hole, there might be a finite 
interval when signals from the singularity can escape to ${\cal I^{+}}\,$
(``cosmic flashing''). However, the analog of Eq. (\ref{eq:Qineq}) for the 
four-dimensional Reissner-Nordstr{\o}m spacetime leads to the constraint
\begin{equation}
|\Delta M| < {1\over t},
\end{equation}
where $\Delta M$ is the change in the black hole's mass due to absorption of
negative energy, and $t$ is the duration of the naked singularity. 
This implies \cite{FR90,FR92}
that the change in the background geometry is less than the expected quantum 
metric fluctuations on the timescale $t$. Thus it is doubtful that the 
naked singularity is observable.

We may also use the quantum inequalities of the form of Eq. (\ref{eq:Qineq}) 
to limit
any violations of the second law of thermodynamics due to negative energy.
Let us consider the use of negative energy to decrease the mass of a black hole
by $\Delta M$. This decreases the entropy of the black hole by an amount
of order
\begin{equation}
\Delta S \approx M \Delta M.
\end{equation}
This entropy decrease can only be sustained for a period of time 
$T \leq (\Delta M)^{-1}$. If we wish to be able to measure the area of 
the horizon and verify that it has decreased, we should require that
this time be larger than the light travel time across the black hole,
so $T >M$. These conditions together imply that
\begin{equation}
\Delta S < 1.
\end{equation}
This small entropy corresponds to less than one bit of information.
Thus, it is clear that negative energy constrained by an inequality
of the form of Eq. (\ref{eq:Qineq}) cannot produce a macroscopic violation 
of the second law. 

It should be noted that there are examples of negative energy fluxes that
do not obey an inequality of this form. An example is the case of an observer
at a fixed value of $r$ just outside of the horizon of an evaporating black
hole. Such an observer sees a constant negative flux going into the hole
on a timescale of the order of the black hole's lifetime. However, this
observer is non-inertial. If one wants to describe measurements made by
a detector carried by such an observer, one needs to take into account the 
Unruh radiation effects, by which an accelerated detector responds as
though it were immersed in a thermal bath. In fact, Unruh \cite{Unruh} 
has shown that
for a detector near the black hole horizon, this effect dominates any effect
due to the ingoing negative energy. There are other apparent counterexamples
to the inequality, Eq. (\ref{eq:Qineq}), which involve inertial motion 
through a negative 
energy background. These include observers moving through the Casimir energy
in a cylinder universe, and observers orbiting an evaporating black 
hole \cite{FR2}.
However, in all of these cases, the quantum field is in its natural ground
state, and it is not clear that it is possible to absorb any of this
negative energy. Furthermore, it is possible to prove ``difference inequalities''
which limit how much more negative one may make the energy density or flux
by changing the quantum state \cite{FR95}.

In addition to the above inequalities on fluxes of negative energy, there
are also  restrictions on negative energy density \cite{FR97}, analogous to
Eqs.~(\ref{eq:flux}) and (\ref{eq:4dflux}). Let
\begin{equation}
 \rho = \langle T_{\mu\nu} \, u^\mu u^\nu \rangle
\end{equation}
be the expectation value of the energy density in the frame of an inertial
observer with four-velocity $u^\mu$, and define the sampled energy density
as
\begin{equation}
\hat \rho \equiv {t_0 \over \pi} \int_{-\infty}^{\infty} 
{{\rho(t) dt}\over {t^2+{t_0}^2}},.   \label{eq:rhohat}
\end{equation} 
For a massless scalar field in two-dimensional flat spacetime, one has that
\begin{equation}
\hat \rho \geq - \frac{1}{8 \pi t_0^2} \,, \label{eq:rhohat2d}
\end{equation}
and in four dimensional flat spacetime, one finds the inequalities
\begin{equation}
\hat \rho \geq - \frac{3}{32 \pi^2 t_0^4}  \label{eq:rhohat4d}
\end{equation}
for the massless scalar field, and
\begin{equation}
\hat \rho \geq - \frac{3}{16 \pi^2 t_0^4}  \label{eq:rhohat4dem}
\end{equation}
for the electromagnetic field. Flanagan \cite{Flanagan} has recently
generalized the two-dimensional inequalities, Eqs.~(\ref{eq:flux}) and
(\ref{eq:rhohat2d}) for arbitrary sampling functions. He has also shown that
the resulting inequalities are optimal in that there exists a quantum state for 
which the inequalities become equalities. The right hand sides of Flanagan's
optimal inequalities differ from those of  Eqs.~(\ref{eq:flux}) and
(\ref{eq:rhohat2d}) by a factor of three.

   In addition to the possible violations of the second law of thermodynamics
or of cosmic censorship discussed above, negative energy can lead to other 
bizarre phenomena. These include ``traversable wormholes'' \cite{MTU}, which are
tunnels which might be used to travel quickly to a distant region of the
universe. Such wormholes would require negative energy densities
in order that their metric be a solution of the semiclassical Einstein
equations. Wormhole solutions have the possibility of containing closed
time-like curves and apparent causality violation. Another spacetime
which requires negative energy is the ``warp drive'' bubble of Alcubierre
\cite{Alcubierre}, which also involves superluminal travel and possible
causality violations \cite{Everett}. Hawking has proposed a ``chronology
protection conjecture'' \cite{Hawking92}, according to which backreaction
from the stress tensor of quantum matter fields might prevent the formation
of closed null or timelike curves. 

     In general, it is much more difficult to prove quantum inequalities in 
curved spacetime than in Minkowski spacetime, and exact results have been
obtained only for a few special cases, such as the Einstein universe \cite{PF97}.
However, so long as the sampling time $t_0$ is small compared to the local 
radius of curvature of spacetime (or the distance to any boundaries), then one
expects the flat space inequalities, such as Eqs.~(\ref{eq:rhohat4d}) and 
(\ref{eq:rhohat4dem}) to be approximately satisfied because spacetime is
approximately Minkowskian on such small scales. This line of reasoning has been
used to place severe constraints on the sizes of traversable wormholes \cite{FR96}
or warp drive geometries \cite{PF97b,ER97}. Note that these restrictions are
independent of the chronology protection conjecture, which would prevent
causality violations, but otherwise place no constraints upon macroscopic
wormhole or warp bubbles.

    Recently, quantum inequalities have been proven in general static curved 
spacetimes \cite{Song,PF97c} in the short sampling time limit. These results
confirm that in this limit one obtains the flat space inequality, plus
subdominant curvature dependent corrections.

\vspace{1cm}

\section{SEMICLASSICAL GRAVITY THEORY \hfil\break AND METRIC FLUCTUATIONS}

\subsection{Limits of the Semiclassical Theory and Stress tensor Fluctuations}
\setcounter{equation}{0}
\renewcommand{\theequation}{5.\arabic{equation}}

     The semiclassical theory of gravity is that in which a classical
gravitational field is coupled to a quantized matter field through the
semiclassical Einstein equations:
\begin{equation}
G_{\mu\nu} =-8 \pi \langle T_{\mu\nu} \rangle. \label{eq:semi}
\end{equation}
This theory provides the necessary transition to the classical theory
of gravity. It also seems to give a convincing picture of the backreaction
to the Hawking radiation. Calculations of $\langle T_{\mu\nu} \rangle$
in the Unruh vacuum state reveal a steady negative energy flux into the
horizon which accounts for the decrease in mass of the black hole as
evaporation proceeds \cite{Elster,JMO}. 

However, this theory also suffers from some serious problems. One of
these is that the semiclassical equations are typically fourth order
equations. The tensors $H^{(1)}_{\mu\nu}$ and $H^{(2)}_{\mu\nu}$, defined
in Eqs. (\ref{eq:H1}) and (\ref{eq:H2}), 
involve fourth derivatives of the metric and will
generally appear as part of $\langle T_{\mu\nu} \rangle$. This can lead 
to runaway solutions \cite{HW,PS}, 
similar to those in classical electron theory when
radiation reaction is taken into account. 

Another difficulty of the theory based upon Eq. (\ref{eq:semi}) is that it fails
when there are large fluctuations in the stress tensor. This may be
illustrated with a simple example: Suppose our system is in a superposition
state in which the two possibilities are a $1000kg$ mass located on either 
side of our laboratory, with equal amplitudes. If we measure the resulting
gravitational field with a gravimeter, we expect to find either the 
gravitational field of $1000kg$ on one side of the laboratory or that of
$1000kg$ on the other side, and that each will occur with $50\%$ probability.
However, Eq. (\ref{eq:semi}) predicts that we will always find 
the gravitational field
produced by having $500kg$ on both sides of the laboratory. This difficulty
is avoided if we only use Eq. (\ref{eq:semi}) when
the quantum state of the system is one in which
the stress tensor fluctuations are small \cite{F82}, that is, one in which
\begin{equation}
\langle T_{\alpha \beta}(x) \,T_{\mu \nu}(y) \rangle \approx
\langle T_{\alpha \beta}(x) \rangle 
\langle T_{\mu \nu}(y) \rangle.   \label{eq:T2}
\end{equation}

Of course, the expectation values on both sides of the above equation are 
formally divergent and need to be defined. Let us restrict our attention
to free fields in Minkowski spacetime. Then all operators will have finite
expectation values if we define them as being normal ordered with respect
to the Minkowski vacuum state. 
Let
\begin{equation}
\Delta(x) \equiv \Biggl| {\langle\colon T_{00}{}^2(x)\colon
\rangle-\langle\colon T_{00}(x)\colon\rangle^2 
\over
\langle\colon T_{00}{}^2(x)\colon\rangle} 
\Biggr|.    \label{eq:Delta} 
\end{equation}
Note that $\langle\colon T_{00}(x)\colon\rangle$ is the mean energy
density at $x$ and $\langle\colon T_{00}{}^2(x)\colon \rangle$ is the
mean squared energy density. Thus $\Delta$ is a measure of the energy
density fluctuations at point $x$. We could define similar quantities
which measure the fluctuations in other stress tensor components.
We should require that $\Delta \ll 1$ in order that the energy density
fluctuations be small and that the semiclassical theory of gravity
be valid. The fluctuations in the other components of $T_{\mu \nu}$
should also be small; however, we will restrict our attention to the
energy density. 

Let us consider a massless, scalar field for which the energy density is 
\begin{equation}
T_{00} = {1\over 2}(\dot \phi^2 +|{\bf \nabla}\phi|^2).
\end{equation}
In a coherent state, one may show that
\begin{equation}
\langle T_{\alpha \beta}(x) \,T_{\mu \nu}(y) \rangle =
\langle T_{\alpha \beta}(x) \rangle 
\langle T_{\mu \nu}(y) \rangle,
\end{equation}
and hence $\Delta = 0$. Thus coherent states  are states of minimum 
stress-energy fluctuations, and are hence states for which the semiclassical
gravity theory holds. This is to be expected as coherent states are the 
quantum states which describe classical field excitations. It is of interest
to now consider squeezed states. As we saw in Lecture 4, this two parameter
family of states includes the coherent states as one limit, but also includes
the squeezed vacuum states with negative energy density as another limit.
 Recall for the general squeezed state
$|z,\zeta \rangle$, that $z$ is the coherent state parameter and $\zeta$
is the squeezing parameter. If $|z| \gg |\zeta|$, then the state is close to
a coherent state, whereas non-classical behavior such as negative energy
densities arise in the opposite limit where $|\zeta| \leq |z|$. 
In Ref. \cite{Kuo}, $\Delta$ was calculated numerically for various 
ranges of these
parameters for a single plane wave mode. It was found that $\Delta \ll 1$ 
holds only in the former limit,
$|z| \gg |\zeta|$. In particular, by the point that $|\zeta|$ has 
increased so that negative energy density appears somewhere, one always
seems to have that $\Delta$ is of order unity. This result implies that
the semiclassical theory of gravity fails for quantum states in which the
energy density is negative.

It is also of interest to compute the fluctuations in the Casimir energy 
density. As noted previously, this can provide an example of negative energy
density. In general, the calculation of the Casimir energy for any but the
simplest geometries is a very difficult task. Nonetheless, it is possible
to establish a lower bound on $\Delta$ which is independent of the boundary
conditions. For the case of a massless scalar field, this lower bound 
is \cite{Kuo}
\begin{equation}
\Delta \geq {1 \over 3}.
\end{equation}
In the particular case of such a field which is periodic in one spatial
direction with periodicity length $L$, the Casimir energy density is
\begin{equation}
\langle\colon T_{00}(x)\colon\rangle
=-{\pi^2\over{90\,L^4}}\, ,
\end{equation} 
and $\Delta = 6/7$. In all cases, $\Delta$ is at least of order unity,
so there are large energy density fluctuations.

Thus our criterion, Eq. (\ref{eq:T2}), for the validity of the semiclassical
gravity theory is not fulfilled for the Casimir energy. This brings us
to the question of how {\it do } we describe the gravitational field 
of the Casimir vacuum. The answer is presumably that we must introduce a 
fluctuating metric, rather than a fixed classical metric. The concept
of a fluctuating metric is perhaps best approached in an operational
manner. We can think of a classical metric as encoding information
about the trajectories of classical test particles. Similarly, a fluctuating
metric may be described in terms of the statistical properties of an ensemble
of test particles. In this way we are led to treat the fluctuating 
gravitational field in terms of the Brownian motion which it produces in test 
particles. 

Brownian motion may be described by means of a Langevin equation. In the
case of nonrelativistic motion on a nearly flat background, this equation
is
\begin{equation}
m{d{\bf v}(x)\over dt}= {\bf F}_c(x) + {\bf F}(x),
\end{equation}
where $m$ is the test particle mass, ${\bf F}_c$ is a classical force,
and ${\bf F}$ is a fluctuating force. In our case, the latter will be the force
produced by the fluctuating gravitational field. 
The solution of this equation is
\begin{equation}
{\bf v}(t)= {\bf v}(t_0) + {1\over m}\int_{t_0}^t 
  [{\bf F}_c(t') + {\bf F}(t')] \,dt'
= {\bf v}_c (t) + {1\over m}\int_{t_0}^t {\bf F}(t') \,dt'\, ,
\end{equation}
where ${\bf v}_c (t)$ is the velocity along a classical trajectory. 
We assume that the fluctuating force averages to zero, 
$\langle {\bf F}\rangle =0$, so $\langle {\bf v}\rangle = {\bf v}_c (t)$,
but that quantities quadratic in ${\bf F}$ do not average to zero. 
Thus the mean squared velocity, averaged over an ensemble of
test particles is, 
\begin{equation}
\langle {\bf v}^2\rangle = {\bf v}^2(t_c) + {1\over m^2}
\int_{t_0}^t dt_1 \int_{t_0}^t dt_2\, 
\langle {\bf F}(t_1)\,{\bf F}(t_2)\rangle\, . \label{eq:vsquare}
\end{equation}

Typically, the correlation function for a fluctuating force vanishes
for times separated by much more than some correlation time, $t_c$,
and is approximately constant for shorter time separations:
\begin{equation}
\langle {\bf F}(t_1)\,{\bf F}(t_2)\rangle \approx
\cases{\langle  F^2 \rangle, & $|t_1-t_2| < t_c$, \cr
       0, & $|t_1-t_2| > t_c$. \cr }
\end{equation}
In this case, the contribution of the fluctuating force to 
$\langle {\bf v}^2\rangle$ grows linearly in time:
\begin{equation}
\langle {\bf v}^2\rangle \sim {\bf v}^2(t_c) + 
        {1\over m^2}\langle  F^2 \rangle\, t_c t \, ,\qquad t \gg t_c \,.
\end{equation}

We can apply these notions to the case of the gravitational field of 
the Casimir vacuum by considering test particles which interact only
gravitationally (that is, have no coupling to the quantized field itself).
In the absence of fluctuations, such a classical particle shot down
parallel to and midway in between a pair of conducting plates would
follow a trajectory half way between the plates indefinitely. However,
the fluctuations of the gravitational field will cause it to eventually
drift toward one plate or the other. The characteristic time scale for 
the fluctuations, $t_c$, will in this case be of order $L$, the plate 
separation. More generally, when the semiclassical theory of 
Eq. (\ref{eq:semi})
breaks down because of large fluctuations in the stress tensor, we
are forced to replace the notion of a classical gravitational field
by a statistical description. (Such a statistical viewpoint has been
investigated in recent years by Hu and coworkers \cite{Hu}.)

\subsection{Metric Fluctuations in Linearized Quantum Gravity}

The metric fluctuations which we have been discussing arise strictly from
fluctuations in the source of the gravitational field, and hence might be
dubbed ``passive'' fluctuations. There can also be fluctuations of the 
gravitational degrees of freedom themselves, which we might call ``active''
fluctuations. The latter will arise from the quantum nature of gravity
and should become important at the Planck scale. In the absence of a full
quantum theory of gravity which is capable of treating the Planck scale,
we may still discuss active fluctuations in the weak field limit. Consider
quantized metric perturbations (gravitons) propagating on a fixed background
spacetime. These gravitons could be in a quantum state, such as a squeezed
vacuum state, in which there are significant fluctuations. For example,
gravitons created in the early universe are expected to be in a squeezed
vacuum state \cite{Grishchuk}. 
On a Robertson-Walker background, they will produce 
Weyl curvature fluctuations in the sense that 
$\langle C_{\alpha\beta\sigma\rho} \rangle = 0$, but
$\langle C_{\alpha\beta\sigma\rho}C^{\alpha\beta\sigma\rho} \rangle \not= 0$.

 Let us consider a flat background spacetime with a linearized perturbation 
$h_{\mu\nu}$  propagating upon it.
In the unperturbed spacetime, the square of the geodesic separation of
points $x$ and $x'$ is $2\sigma_0 =(x-x')^2$. In the presence of the
perturbation, let this squared separation be $2\sigma$, and write
\begin{equation}
\sigma= \sigma_0 + \sigma_1 + O(h_{\mu\nu}^2),
\end{equation}
so $\sigma_1$ is the first order shift in $\sigma$. 

In flat spacetime, the retarded Green's function for a massless scalar field is
\begin{equation}
G_{ret}^{(0)}(x-x') = {{\theta(t-t')}\over {4\pi}} \delta(\sigma_0)\, ,
\end{equation}
which has a delta-function singularity on the future lightcone and is zero
elsewhere. In the presence of a classical
metric perturbation, the retarded Green's function has its delta-function
singularity on the perturbed lightcone, where $\sigma=0$. In general,
it may also become nonzero on the interior of the lightcone due to 
backscattering off of the curvature. However, we are primarily interested
in the behavior near the new lightcone, and so let us replace 
$G_{ret}^{(0)}(x-x')$ by 
\begin{equation}
G_{ret}(x,x') = {{\theta(t-t')}\over {4\pi}} \delta(\sigma)\, .
\end{equation}
This may be expressed as
\begin{equation}
G_{ret}(x,x') = {{\theta(t-t')}\over {8\pi^2}} \int_{-\infty}^{\infty}
                d\alpha\, e^{i\alpha \sigma_0}\, e^{i\alpha \sigma_1}\, .
\end{equation}

We now replace the classical metric perturbations by gravitons in a squeezed
vacuum state $|\psi\rangle$. Then $\sigma_1$ becomes a quantum operator 
which is linear
in the graviton field operator, $h_{\mu\nu}$. Because a squeezed vacuum state
is a state such that $\sigma_1$ may be decomposed into positive and negative
frequency parts, i.e., we may find $\sigma_1^{+}$ and $\sigma_1^{-}$ so that
$\sigma_1^{+} |\psi\rangle =0$, $\langle \psi| \sigma_1^{-}=0$, and 
$\sigma_1 = \sigma_1^{+} + \sigma_1^{-}$. Thus, the derivation of 
Eq. (\ref{eq:eiphi})
holds here as well and enables us to write
\begin{equation}
\Bigl\langle e^{i\alpha \sigma_1} \Bigr\rangle = 
e^{-{1\over 2}\alpha^2 \langle \sigma_1^2 \rangle} \, .
\end{equation}
Thus when we average over the metric fluctuations, the retarded Green's
function is replaced by its quantum expectation value:
\begin{equation}
\Bigl\langle G_{ret}(x,x') \Bigr\rangle = 
{{\theta(t-t')}\over {8\pi^2}} \int_{-\infty}^{\infty}
                                       d\alpha \, e^{i\alpha \sigma_0} \,
e^{-{1\over 2}\alpha^2 \langle \sigma_1^2 \rangle} \, .
\end{equation}
This integral converges only if $\langle \sigma_1^2 \rangle > 0$, in which
case it may be evaluated to yield
\begin{equation}
\Bigl\langle G_{ret}(x,x') \Bigr\rangle = 
{{\theta(t-t')}\over {8\pi^2}} \sqrt{\pi \over {2\langle \sigma_1^2 \rangle}}
\; \exp\Bigl(-{{\sigma_0^2}\over {2\langle \sigma_1^2 \rangle}}\Bigr)\, .
                                           \label{eq:retave}
\end{equation}
Note that this averaged Green's function is indeed finite at $\sigma_0 =0$
provided that $\langle \sigma_1^2 \rangle \not= 0$. Thus the lightcone 
singularity has been smeared out. Note that the smearing occurs in 
{\it both} the timelike and spacelike directions. This is illustrated in
Fig. 5. 

\begin{figure}
\begin{center}
\leavevmode\epsfysize=6cm\epsffile{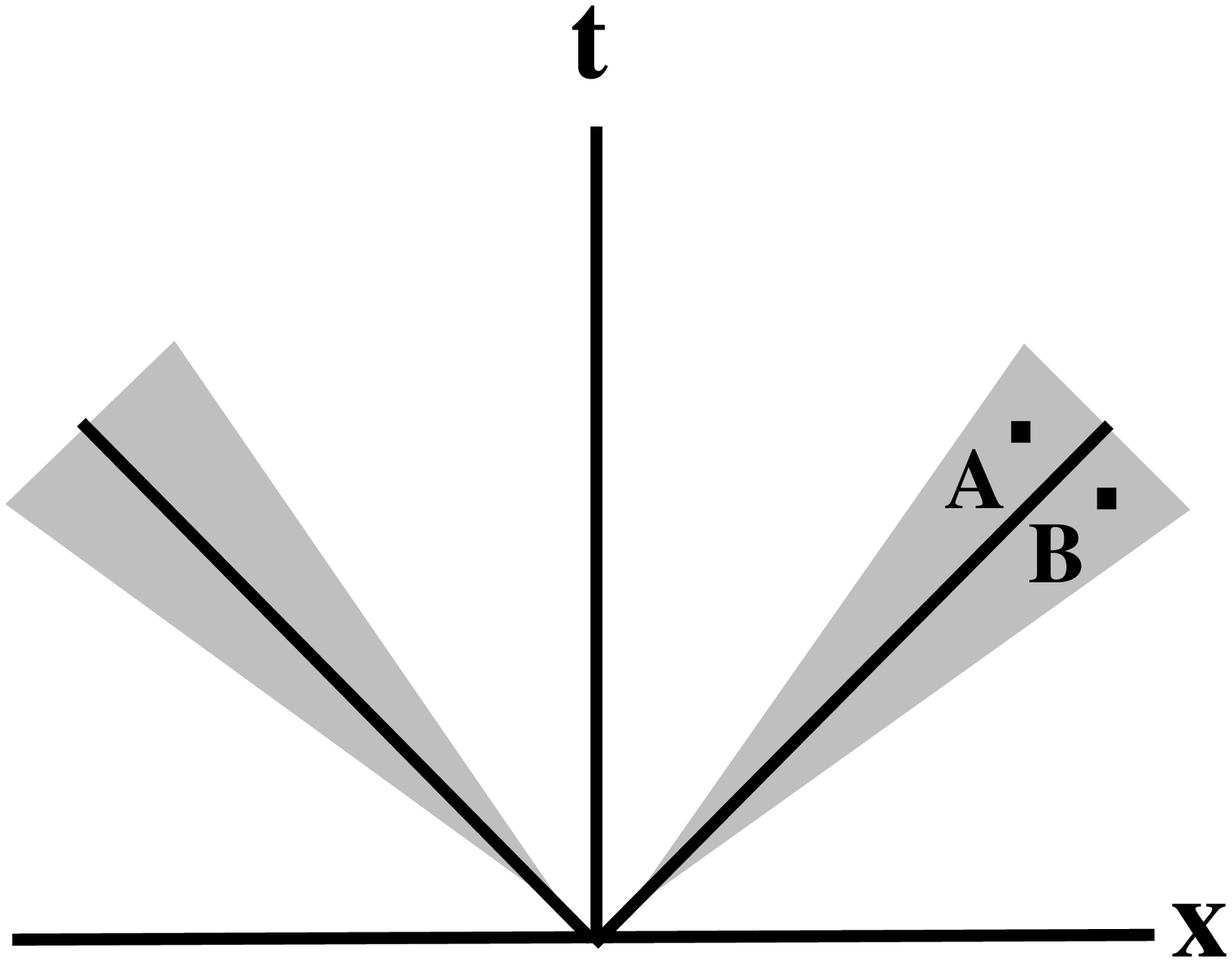}
\label{Figure 5}
\end{center}
\begin{caption}[]

The smearing of the lightcone due to metric fluctuations.
A photon which arrives at Point A from the origin has been slowed by the
effect of metric fluctuations. A photon which arrives at Point B has been
boosted by metric fluctuations, and appears to travel at a superluminal
velocity in the background metric.   
\end{caption}
\end{figure}

This smearing may be interpreted as due to the fact that photons may be
either slowed down or boosted by the metric fluctuations. Photon propagation
now becomes a statistical phenomenon; some photons travel slower than 
light on the classical spacetime, whereas others travel faster. We have now
the possibility of ``faster than light'' signals. This need not cause any
causal paradoxes, however, because the theory is no longer Lorentz invariant.
The graviton state defines a preferred frame of reference. The mean deviation 
from the classical propagation time is 
\begin{equation}
\Delta t = \frac{\sqrt{\langle \sigma_1^2 \rangle}}{r} \,, \label{eq:Dt}
\end{equation}
where $r$ is the separation between the source and the detector. As expected,
this characteristic time delay or advance depends upon the density of gravitons
present through the quantity $\langle \sigma_1^2 \rangle$. It should be
noted that $\Delta t$ is an ensemble averaged quantity, the characteristic time 
delay or advance averaged over many trials. If the source emits a pair of
photons in rapid succession, the expected difference in the flight times of
the two photons is typically less than $\Delta t$. This arises because each
photon probes a classical geometry which is almost the same \cite{FS96}.

   This discussion of lightcone fluctuations may be generalized to curved 
spacetime \cite{FS96,FS97}. When applied to classical spacetimes which possess 
horizons, one has {\it horizon fluctuations}. This phrase is perhaps an
oxymoron, as a fluctuating horizon is no longer a horizon at all in the 
classical sense. Rather than a precise boundary determining which regions
may not communicate with the outside world, one now has the possibility
(however small) that information can leak across classical horizons.

In a similar way to the above calculation of the averaged retarded Green's 
function, we may average the other singular functions over metric
fluctuations. For example, for the case that $\langle \sigma_1^2 \rangle > 0$, 
the Hadamard function becomes
\begin{equation}
\Bigl\langle G_{1}(x,x') \Bigr\rangle = -{1 \over{2\pi^2}}\Bigl\langle
    {1\over \sigma} \Bigr\rangle = -{1 \over{2\pi^2}}
\int_{0}^{\infty} d\alpha\: \sin\,{\alpha \sigma_0} \:
e^{-{1\over 2}\alpha^2 \langle \sigma_1^2 \rangle} \, .
\end{equation}
In the limit that $\sigma_0^2 \gg \langle \sigma_1^2 \rangle$, we recover
the usual form of $G_{1}$ :
\begin{equation}
\Bigl\langle G_{1}(x,x') \Bigr\rangle \sim -{1 \over{2\pi^2}} {1\over \sigma_0}.
\end{equation}
On the other hand, near the lightcone, $\Bigl\langle G_{1}(x,x') \Bigr\rangle$
is finite:
\begin{equation}
 \Bigl\langle G_{1}(x,x') \Bigr\rangle \sim 
-{{\sigma_0} \over{2\pi^2 \langle \sigma_1^2 \rangle}}, \qquad
  \sigma_0^2 \ll \langle \sigma_1^2 \rangle.    \label{eq:Hadave}
\end{equation}

The average of the Feynman propagator over metric fluctuations may be obtained
from Eqs. (\ref{eq:retave}) and (\ref{eq:Hadave}) and the relation
\begin{equation}
G_{F}(x,x')= {1\over 2} [G_{ret}(x,x') +G_{ret}(x',x)] -iG_{1}(x,x'),
\end{equation}
and is also finite on the lightcone. 

These averaged functions are, however, not finite in the limit of coincident 
points, that is in the limit that both $\sigma_0^2$ and 
$\langle \sigma_1^2 \rangle$ vanish with their ratio finite. This can be 
understood on the grounds that the effect of metric fluctuations is to cause
the propagation time for a photon to fluctuate. This causes an effect which
grows with increasing spatial separation, but is small for points which
are spatially close to one another.

  It was conjectured long ago \cite{Pauli,Deser,DeWitt} that quantum metric
fluctuations might smear out the singularities of Green's functions on the 
light cone, and thereby solve the ultraviolet divergence problems of quantum
field theory. The model described in this  lecture does smear out the 
lightcone singularities, but it does not remove all of the 
ultraviolet divergences of quantum field theory.

\vspace{0.5cm}

{\bf Acknowledgement:} This work was supported in part by the National
Science Foundation under Grant PHY-9507351.

\end{document}